\newcommand{\sx}{\sigma_x}

\newcommand{\sz}{\sigma_z}

\newcommand{\SL}{S_{\textnormal{L}}}
\newcommand{\SR}{S_{\textnormal{R}}}
\newcommand{\TR}{T_{\textnormal{R}}}
\newcommand{\TL}{T_{\textnormal{L}}}
\newcommand{\mb}{\mu_{\textnormal{B}}}
\newcommand{\pss}{P_{\textnormal{SS}}}
\newcommand{\pst}{P_{\textnormal{ST}}}
\newcommand{\pts}{P_{\textnormal{TS}}}
\newcommand{\ptt}{P_{\textnormal{TT}}}
\newcommand{\RL}{r_{\textnormal{L}}}
\newcommand{\RR}{r_{\textnormal{R}}}
\newcommand{\GL}{g_{\textnormal{L}}}
\newcommand{\GR}{g_{\textnormal{L}}}

\newcommand{\tn}[1]{\textnormal{#1}}
\newcommand{\B}{\textnormal{B}}
\newcommand{\U}{\uparrow}
\newcommand{\D}{\downarrow}

\documentclass[%
reprint,amsmath,amssymb,aps,prl,superscriptaddress,longbibliography,
]{revtex4-1}

\usepackage{graphicx}
\usepackage{dcolumn}
\usepackage{bm}
\usepackage{epstopdf}
\usepackage{float}
\usepackage{braket}
\usepackage{dsfont}
\usepackage{lipsum}
\usepackage{xr}

\begin{document}

\title{Conditional Teleportation of Quantum-Dot Spin States}

\author{Haifeng Qiao}
\thanks{These authors contributed equally.}

\author{Yadav P. Kandel}
\thanks{These authors contributed equally.}

\affiliation{Department of Physics and Astronomy, University of Rochester, Rochester, NY, 14627 USA}

\author{Sreenath K. Manikandan}
\affiliation{Department of Physics and Astronomy, University of Rochester, Rochester, NY, 14627 USA}

\author{Andrew N. Jordan}
\affiliation{Department of Physics and Astronomy, University of Rochester, Rochester, NY, 14627 USA}
\affiliation{Institute for Quantum Studies, Chapman University, Orange, CA 92866, USA}

\author{Saeed Fallahi}
\affiliation{Department of Physics and Astronomy, Purdue University, West Lafayette, IN, 47907 USA}
\affiliation{Birck Nanotechnology Center, Purdue University, West Lafayette, IN, 47907 USA}

\author{Geoffrey C. Gardner}
\affiliation{Birck Nanotechnology Center, Purdue University, West Lafayette, IN, 47907 USA}
\affiliation{School of Materials Engineering, Purdue University, West Lafayette, IN, 47907 USA}

\author{Michael J. Manfra}
\affiliation{Department of Physics and Astronomy, Purdue University, West Lafayette, IN, 47907 USA}
\affiliation{Birck Nanotechnology Center, Purdue University, West Lafayette, IN, 47907 USA}
\affiliation{School of Materials Engineering, Purdue University, West Lafayette, IN, 47907 USA}
\affiliation{School of Electrical and Computer Engineering, Purdue University, West Lafayette, IN, 47907 USA}

\author{John M. Nichol}
\email{john.nichol@ur.rochester.edu}

\affiliation{Department of Physics and Astronomy, University of Rochester, Rochester, NY, 14627 USA}

\begin{abstract}
Among the different platforms for quantum information processing, individual electron spins in semiconductor quantum dots stand out for their long coherence times and potential for scalable fabrication. The past years have witnessed substantial progress in the capabilities of spin qubits. However, coupling between distant electron spins, which is required for quantum error correction, presents a challenge, and this goal remains the focus of intense research. Quantum teleportation is a canonical method to transmit qubit states, but it has not been implemented in quantum-dot spin qubits. Here, we present evidence for quantum teleportation of electron spin qubits in semiconductor quantum dots. Although we have not performed quantum state tomography to definitively assess the teleportation fidelity, our data are consistent with conditional teleportation of spin eigenstates, entanglement swapping, and gate teleportation. Such evidence for all-matter spin-state teleportation underscores the capabilities of exchange-coupled spin qubits for quantum-information transfer.
\end{abstract}



\maketitle
\section{Introduction}
Quantum teleportation~\cite{Bennett1993} is an exquisite example of the power of quantum information transfer. Teleportation has been demonstrated in many experimental quantum information processing  platforms~\cite{Bouwmeester1997,Riebe2004,Olmschenk2009,Steffen2013,Pfaff2014,Pirandola2015}, and it is an essential tool for quantum error correction~\cite{Knill2005}, measurement-based quantum computing~\cite{Raussendorf2003}, and quantum gate teleportation~\cite{Gottesman1999}. However, quantum teleportation has not previously been demonstrated in quantum-dot spin qubits. Separating entangled pairs of spins to remote locations, as required for quantum teleportation, has previously presented the main challenge to teleportation in quantum dots. 

\begin{figure}
	\includegraphics{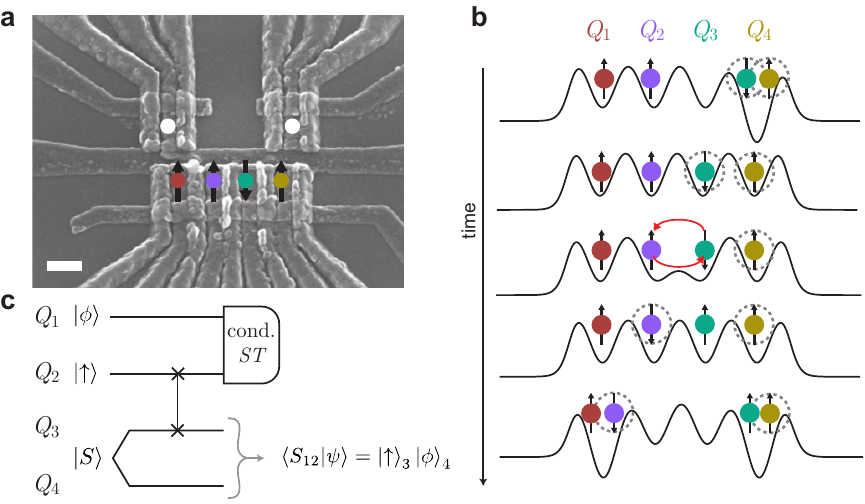}
	\caption{\label{apparatus} Experimental setup. (a) Scanning electron micrograph of the quadruple quantum dot. The positions of the electron-spin qubits are overlaid. The white dots indicate the positions of the sensor quantum dots. The scale bar is 200 nm. (b) Physical implementation of the teleportation protocol. Dots 3 and 4 are initialized in the singlet configuration via electron exchange with the reservoirs and then separated via tunneling. We implement the SWAP gate as a positive voltage pulse to the barrier gate between dots 2 and 3. Pairs of qubits are measured in the singlet/triplet basis via Pauli spin blockade. (c) Circuit diagram for the conditional quantum teleportation protocol. $\ket{\psi}$ represents the four-qubit wavefunction.}
\end{figure}

Here, we overcome this challenge using a recently demonstrated technique to distribute entangled spin states via Heisenberg exchange~\cite{Kandel2019}. This technique does not involve the motion of electrons, greatly simplifying the teleportation procedure. Our teleportation method also leverages Pauli spin blockade, a unique feature of electrons in quantum dots, to generate and measure entangled pairs of spins. We combine these concepts to perform conditional teleportation in a system of four GaAs quantum-dot spin qubits.  Our data are consistent with conditional teleportation of quantum-dot spin states, entanglement swapping, and gate teleportation. Entanglement swapping~\cite{Zukowskwi1993} goes beyond teleportation of single-qubit states to create entanglement between uncorrelated particles via measurements, and demonstrations of entanglement swapping in matter qubits are rare~\cite{Riebe2008,Ning2019}. Our technique is fully compatible with all gate-defined quantum-dot types, including Si quantum dots. Although we use coherent spin-state transfer via Heinseberg exchange~\cite{Kandel2019} to distribute entangled pairs of spins, other methods to create long-range-entangled states of spins including tunneling~\cite{Fujita2017,Nakajima2018,Flentje2017} and coupling via superconducting resonators~\cite{Borjans2020} could be used as well.

\begin{figure*}
	\includegraphics{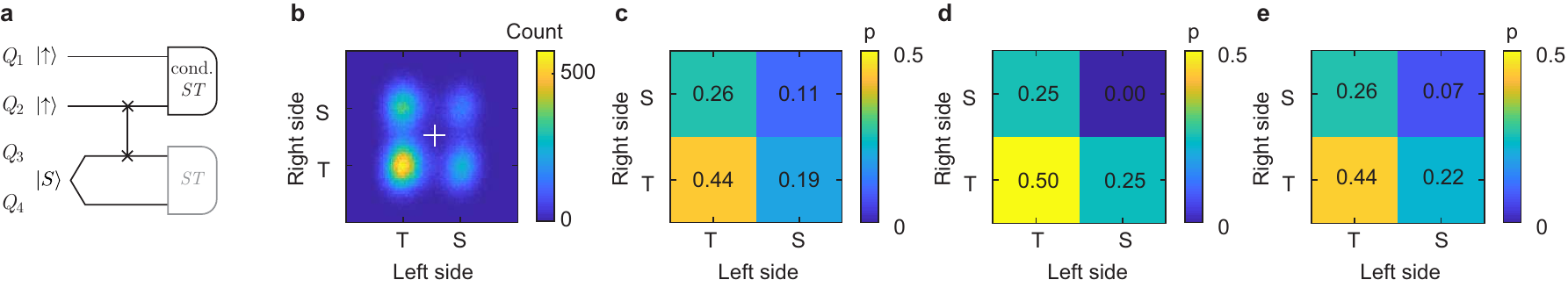}
	\caption{\label{teleport} Conditional teleportation of a classical spin state. (a) Quantum circuit to teleport a state $\ket{\U}$ from qubit 1 to qubit 4. The conditional singlet-triplet measurement on qubits 1 and 2 induces teleportation, and the gray singlet-triplet measurement of the right pair verifies teleportation. (b) Experimentally measured probability distribution for $65{,}536$ single-shot realizations of the teleportation sequence in (a). The white cross indicates the threshold used to calculate probabilities. (c) Extracted probabilities $p$ from the distribution in (b). (d) Simulated probabilities computed neglecting any errors. (e) Simulated probabilities accounting for readout errors, state preparation errors, charge noise, and hyperfine fields. All probabilities are rounded to the nearest hundredth.}
\end{figure*}

\section{Results}
\subsection{Device description}
We implement our teleportation method in a four-qubit quantum processor, which consists of a quadruple quantum dot fabricated in a GaAs/AlGaAs heterostructure  [Fig.~\ref{apparatus} (a)]. Because the ground state wavefunction of two electrons has the spin-singlet configuration, initialization of two spins in a single quantum dot automatically generates an entangled pair of spins~\cite{Petta2005,Foletti2009}. Furthermore, spin-to-charge conversion via Pauli spin blockade~\cite{Petta2005,Barthel2009} enables rapid single-shot measurement of pairs of electron spins in the  $\{ \ket{S}, \ket{T}\}$ basis, where $|T\rangle$ is any one of the triplet states $\{\ket{{\U\U}},\frac{1}{\sqrt{2}} \left(\ket{\U \D}+\ket{\D \U} \right), \ket{\D \D}\}$. We therefore configure the quadruple quantum dot as two pairs of spins to facilitate teleportation. Spins 1 and 2 form the left pair, and spins 3 and 4 form the right pair.

We achieve separation and distribution of entangled pairs of spins through coherent spin-state transfer based on Heisenberg exchange~\cite{Kandel2019}. To transfer a spin state from one electron to another, we induce exchange coupling between electrons by applying a voltage pulse to the barrier gate between them [Fig.~\ref{apparatus}(b)] ~\cite{Martins2016,Reed2016}. Because exchange coupling generates a SWAP operation, this procedure interchanges the two states. This procedure can be repeated for different pairs of spins to enable long-distance spin-state transfer. Importantly, exchange-based spin swaps preserve entangled states~\cite{Kandel2019}. 

\subsection{Conditional teleportation protocol}
Figure~\ref{apparatus}(c) shows the quantum circuit for our procedure, which can conditionally teleport an arbitrary state $\ket{\phi}$ from dot 1 to dot 4. We prepare qubit 2 in the $\ket{\U}$ state, and it is used later for readout, as discussed further below. We generate the Einstein-Podolsky-Rosen (EPR) pair between qubits 3 and 4 by loading two electrons into the right-most dot via electrical exchange with reservoirs.  We then separate the two electrons via tunneling. After a SWAP gate on qubits 2 and 3, the EPR pair resides in qubits 2 and 4. To teleport $\ket{\phi}$ from qubit 1 to qubit 4, we project the left pair of qubits onto the $\{\ket{S},\ket{T}\}$ basis via diabatic charge transfer into the outer dots~\cite{Kandel2019} [Fig.~\ref{apparatus}(b)]. Our measurements in the $\{ \ket{S},\ket{T} \}$ basis can only distinguish $\ket{S}=\ket{\Psi^-}$ from the other Bell states $\ket{\Psi^+}$,  $\ket{\Phi^+}$, or  $\ket{\Phi^-}$, which are linear combinations of the triplet states. In this case, therefore, successful teleportation requires obtaining a singlet in the left pair. To verify teleportation, we also project the right pair, using either diabatic or adiabatic charge transfer (see Methods). 

The utility of quantum teleportation lies in its ability to transmit unknown quantum states. Usually, teleportation of unknown states is experimentally demonstrated by verifying teleportation of a complete set of single-qubit basis states~\cite{Bouwmeester1997} or through process tomography~\cite{Steffen2013}. Because our four-qubit device does not incorporate a micromagnet or antenna for magnetic resonance, we are not able to prepare superposition states of single spins. Therefore, to illustrate the operation of the teleportation procedure, we first teleport a classical spin state from qubit 1 to qubit 4. Later, we demonstrate entanglement swapping in our four-qubit processor, which conclusively demonstrates non-local manipulation of quantum states via measurements. In the future, quantum state tomography will be required to establish that the teleportation fidelity exceeds the classical bound, as discussed below.

To demonstrate the basic operation of our teleportation method using $\ket{\phi}=\ket{\U}$, we prepare qubits 3 and 4 in a spin singlet [Fig.~\ref{teleport}(a)].  Qubits 1 and 2 are prepared in the $\ket{\psi_{12}}=\ket{\phi}_1\ket{\U}_2=\ket{\U}_1\ket{\U}_2$ state by electrical exchange with the reservoirs (see Methods). After the SWAP operation, if the left pair projects onto $\ket{S_{12}}$, qubit 4 should be identically $\ket{\U}$~\cite{Bennett1993}. Because qubit 3 has the $\ket{\U}$ state (a result of the earlier SWAP operation), the right pair should be in the $\ket{\psi_{34}}=\ket{\U}_3\ket{\phi}_4=\ket{\U}_3\ket{\U}_4$ state, and measuring a singlet on the left pair should perfectly correlate with measuring a triplet on the right pair. Figure~\ref{teleport}(b) displays a joint histogram of $65{,}536$ single-shot measurements on both pairs of qubits for the teleport operation discussed above. Figure~\ref{teleport}(c) shows the extracted probabilities for the different outcomes. Our measurements closely match the predicted probabilities, as shown in Fig.~\ref{teleport}(d) (see Methods). Figure~\ref{teleport}(e) shows a prediction including known sources of experimental error, including readout fidelity, relaxation during readout, state preparation error, charge noise, and hyperfine fields, and this prediction matches the observed data closely. We discuss these errors further below. We have also performed similar experiments with qubit 1 prepared in a mixed state (Supplementary Fig.~1), and the results are consistent with our expectations. 

\begin{figure}
	\includegraphics{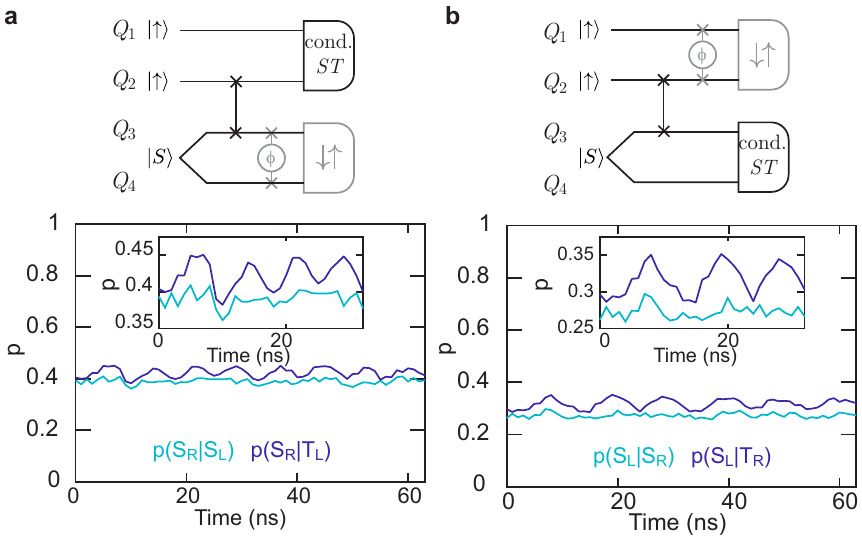}
	\caption{\label{exchange} Verification of conditional teleportation of a classical state. (a) We apply an exchange pulse to qubits 3 and 4 after measuring qubits 1 and 2. Here, $\phi=2\pi J_{34}t$, where $J_{34}$ is the induced exchange coupling between qubits 3 and 4, and $t$ is the evolution time given by the $x$-coordinate of each data point. ($\phi=\pi$ corresponds to a SWAP operation.)  When the left pair give a singlet, the right pair have the same spin, and no oscillations should be visible. The inset shows the same data from 0-32 ns. (b) Applying an exchange pulse to the left pair after measuring the right pair generates exchange oscillations on the left pair only if the right pair yields a triplet. The inset shows the same data from 0-32 ns. Here, $\phi=2\pi J_{12}t$, where $J_{12}$ is the induced exchange coupling between qubits 1 and 2. In both panels, each data point represents the average of $16{,}384$ single-shot measurements, and the gray elements of the circuits serve to verify teleportation.}
\end{figure}

To verify conditional teleportation of the classical state, we perform an exchange gate on qubits 3 and 4 following the teleport [Fig.~\ref{exchange}(a)]. In the case of successful teleportation, qubits 3 and 4 should have the $\ket{\psi_{34}}=\ket{\U}_3\ket{\U}_4$ state, and the exchange gate should have no effect. Indeed, after measuring a singlet on the left pair, we do not observe significant exchange oscillations on the right pair, but after measuring a triplet on the left pair, we do observe exchange oscillations on the right pair [Fig.~\ref{exchange}(a)]. As shown in Supplementary Fig.~2, eliminating the SWAP operation between qubits 2 and 3 or preparing a product state, instead of an EPR pair, on the right side, largely eliminate the conditional effect, consistent with our simulations (Supplementary Fig.~3). These data demonstrate that the both the EPR pair and the SWAP operation are critical for teleportation, as expected. 

The circuit of Fig.~\ref{teleport}(a) can also teleport the state of qubit 3 to qubit 2, depending on the result of the right-pair measurement. To verify that teleportation can also occur from qubit 3 to qubit 2, we switched the order of measurements and performed the variable exchange gate on the left pair of qubits [Fig.~\ref{exchange}(b)]. In this case, we observe that the oscillations on the left pair depend on the state of the right pair. Again, removing the SWAP operation or the EPR pair significantly eliminates the conditional effect (Supplementary Fig.~2). We have performed simulations (see Methods) which include known sources of error, that match our observed data closely, as shown in Supplementary Fig.~3. Our simulations reproduce the weak residual oscillations in $p(\SL|\SR)$ and $p(\SR|\SL)$ (Fig.~3), which likely result from an imperfect SWAP operation and readout errors. Supplementary Fig.~4 shows the expected ideal results for these measurements in the absence of any errors.

\begin{figure*}
	\includegraphics{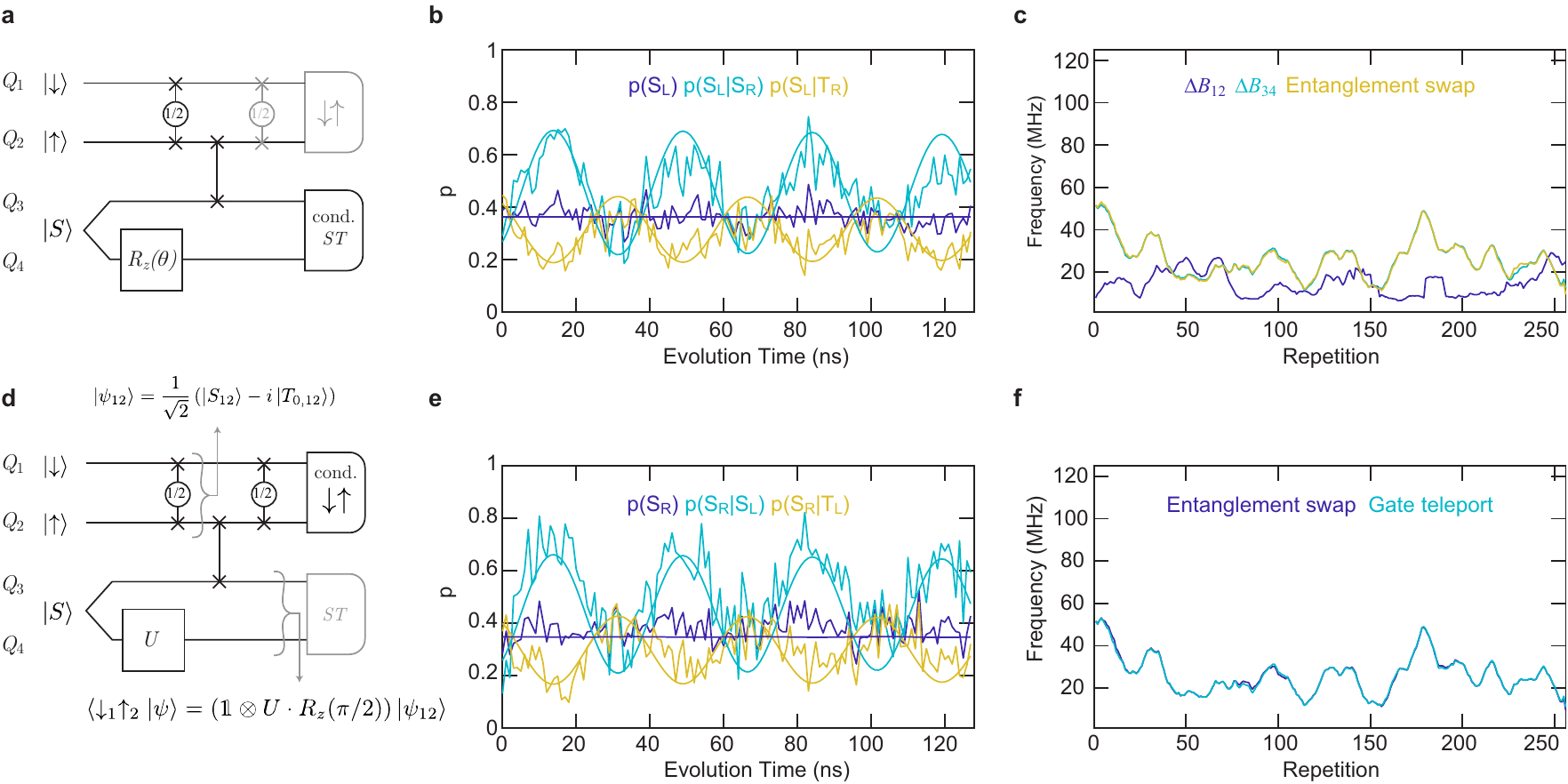}
	\caption{\label{entangle} Conditional entanglement swapping and gate teleportation. (a) Circuit diagram for conditional entanglement swapping. The gray circuit elements are used to verify entanglement swapping. (b) The unconditioned singlet probability on the left side $p(\SL)$ shows no oscillations. However, $p(\SL|\SR)$ and $p(\SL|\TR)$ show pronounced singlet-triplet oscillations. Simulated predictions are shown in the same color. (c) Comparison of the extracted oscillation frequency vs. repetition number measured on qubits 1 and 2 to the measured hyperfine gradients. The sudden jump in $\Delta B_{12}$ near repetition 200 likely results from frequencies too low to measure properly. (d)  Circuit diagram for conditional gate teleportation. (e) $p(\SR)$ shows only weak oscillations, but $p(\SR|\SL)$ and $p(\SR|\TL)$ show prominent singlet-triplet oscillations. Simulated predictions are shown in the same color. (f) Comparison of the extracted entanglement-swap and gate-teleport oscillation frequencies vs. repetition number.}
\end{figure*}

\subsection{Conditional entanglement swapping and gate teleportation}
Having illustrated the basic operation of the teleport procedure, we now present evidence for conditional entanglement swapping, which confirms that the four-qubit processor indeed performs non-local coherent manipulation of quantum information using measurements [Fig.~\ref{entangle}(a)]. Entanglement swapping~\cite{Zukowskwi1993} uses teleportation to generate entanglement between uncorrelated particles via measurements. In this case, we prepare the Einstein-Podolsky-Rosen (EPR) state between qubits 1 and 2 via a $\sqrt{\textnormal{SWAP}}$ gate, starting from the $\ket{\D}_1 \ket{\U}_2$ state. This process generates the entangled state $\frac{1}{\sqrt{2}}\left(\ket{S_{12}}-i\ket{T_{0,12}} \right)$, where $\ket{T_0}=\frac{1}{\sqrt{2}}\left( \ket{\U \D}+\ket{\D \U}\right)$. At the same time, we prepare a separated singlet between qubits 3 and 4. Before teleportation, we evolve the separated singlet in its local hyperfine gradient  $\Delta B_{34}$ for a variable time $t$. This evolution generates an effective $z$-rotation on qubit 4 relative to qubit 3 by an angle $\theta=g \mb\Delta B_{34} t/\hbar$, where $g$ is the electron $g$ factor in GaAs, and $\mb$ is the Bohr magneton. The $z$-rotation on qubit 4 coherently rotates the joint state of qubits 3 and 4 to $\cos(\theta/2+\pi/4)\ket{S}+\exp(- i \pi/2)\sin(\theta/2+\pi/4)\ket{T_0}$~\cite{Petta2005,Foletti2009}. During this evolution time, qubits 3 and 4 remain maximally entangled. 

After a SWAP gate between qubits 2 and 3, qubits 1 and 3 are entangled, and qubits 2 and 4 are entangled.  After projection to a singlet on the right side, entanglements have been swapped, because  the entangled state of qubit 4 is teleported to qubit 1. Qubits 1 and 2, which were not entangled immediately before the measurement, become entangled, provided qubits 3 and 4 project onto the singlet state. Moreover, the coherent singlet-triplet evolution that occurred on qubits 3 and 4 should appear on qubits 1 and 2, given a singlet outcome on the right pair (see Supplementary Note 1).  To verify entanglement swapping, we measure the left pair of qubits by adiabatic charge transfer~\cite{Petta2005,Foletti2009} (see Methods) following another  $\sqrt{\textnormal{SWAP}}$ gate. In the case of successful entanglement swapping, the final $\sqrt{\textnormal{SWAP}}$ gate preserves the coherence of the teleported state against the effects of hyperfine fluctuations during readout.

To observe the anticipated oscillations, we sweep $t$, which controls the $z$ rotation on qubit 4, from 0 to 127 ns, in steps of 1 ns. For each time interval, we implement the quantum circuit shown in Fig.~\ref{entangle}(a) and record a single-shot measurement of both pairs of qubits, and we average this set of measurements 256 times. Figure~\ref{entangle}(b) shows the average of one such set of measurements. No oscillations are visible in the unconditioned singlet probability of the left pair $p(\SL)$. However, prominent oscillations are visible in the probability of a singlet on the left given a singlet on the right $p(\SL|\SR)$ and also in $p(\SL|\TR)$, in good agreement with our simulations [Fig.~\ref{entangle}(b) and Supplementary Fig.~6]. These oscillations demonstrate conditional entanglement swapping. 

Because the nuclear hyperfine fields fluctuate in time, we repeat this set of measurements 256 times, and the entire data set is shown in Supplementary Fig.~7. In between each set, we also perform additional measurements to determine the hyperfine gradients between dots 1 and 2 ($\Delta B_{12}$) and dots 3 and 4 ($\Delta B_{34}$) (Supplementary Fig.~8)~\cite{Foletti2009}. In total, each repetition takes about one second. 

For each repetition, we extract the oscillation frequency by taking a fast Fourier transform of the data (see Methods and Supplementary Fig.~7). Figure~\ref{entangle}(c) shows the extracted oscillation frequency that appears on qubits 1 and 2 after entanglement swapping in addition to the frequencies corresponding to $\Delta B_{12}$ and $\Delta B_{34}$, which were measured concurrently with the teleportation. The observed oscillation frequency measured on qubits 1 and 2 clearly matches the measured hyperfine gradient $\Delta B_{34}$. Because $\Delta B_{12}$ and $\Delta B_{34}$ result from independent nuclear spin ensembles, they evolve differently in time. We note the good agreement between the time evolution of the oscillation frequency after entanglement swapping and the gradient $\Delta B_{34}$.             

To confirm that the singlet-triplet oscillations on the left pair result from entanglement swapping, we have performed additional measurements which omit the SWAP operation between qubits 2 and 3 (Supplementary Fig.~5). These data show no conditional effect. Therefore, the observed oscillations on qubits 1 and 2 in Fig.~\ref{entangle}(b) result entirely from the coherent evolution between entangled states of qubits 3 and 4, together with the SWAP gate and Bell-state measurement. This demonstration of entanglement swapping using our four-qubit processor confirms that we can perform non-local coherent manipulation on entangled states of the form $\cos(\theta/2)\ket{S}+\exp(\pm i \pi/2)\sin(\theta/2)\ket{T_0}$ by quantum measurements. 

A similar circuit [Fig.~\ref{entangle}(d]) also implements a simple example of conditional quantum gate teleportation~\cite{Gottesman1999}, provided that we post-select on the left-side measurements, instead of the right side. In this case, the EPR pair initially consists of qubits 3 and 4, and we teleport qubit 1 to qubit 4. A unitary gate $U$ (the same $z$ rotation discussed above), which is applied to one member of the EPR pair before teleportation, appears on qubit 4 after teleportation. The initial entangled state of qubits 1 and 2 is $\ket{\psi_{12}}=\frac{1}{\sqrt{2}}\left( \ket{S_{12}}-i\ket{T_{0,12}}\right)$. Following the SWAP and conditional teleportation of qubit 1 to qubit 4, qubits 3 and 4 have the state $\ket{\psi_{34}}=(\mathds{1} \otimes U \cdot R_z(\pi/2)) \ket{\psi_{12}}$, and $U$ has been applied to qubit 4. The added $z$ rotation on qubit 4 occurs because of the additional $\sqrt{\textnormal{SWAP}}$ and measurement via adiabatic charge transfer on the left side [Fig.~\ref{entangle}(d)]. 

We measure the right pair of qubits via diabatic charge transfer to verify teleportation [Fig.~\ref{entangle}(e)]. The unconditioned data show very weak oscillations, likely due to an imperfect SWAP gate~\cite{Kandel2019}. Post-selecting based on singlet outcomes on the left side yields prominent oscillations in time, consistent with our simulations. The extracted oscillation frequency versus repetition number agrees well with the data from Fig.~\ref{entangle}(c), as shown in Fig.~\ref{entangle}(f). 

The fidelity of the teleport operation is limited by readout fidelity, relaxation during readout, state preparation, charge noise, and the hyperfine coupling between the electron spins and Ga and As nuclear spins in the substrate. Readout fidelity and relaxation both limit the probability that we will correctly measure the Bell state of one of the EPR pair and the qubit to be teleported. Readout fidelities are $0.93$ for the left pair and $0.87$ for the right pair (see Methods and Supplementary Fig.~11 and Supplementary Fig.~12). State preparation of the EPR pair also affects the teleport operation. We estimate the probability that we correctly prepare the singlet state in dots 3-4 is $0.89$, based on our experimental characterization of the loading process [Supplementary Fig.~10(b)]. Charge noise causes dephasing of the SWAP operation, and the nuclear hyperfine field limits the fidelity of the SWAP operation that we use to transmit the entangled pair of electrons~\cite{Kandel2019}. The simulations shown in Figs.~\ref{teleport} and~\ref{entangle} and Supplementary Figs.~1, 3, and 5 include all of these effects, in addition to the classical-state initialization error (see Methods) where appropriate.

To assess the fidelity of the teleport operation itself for classical states, we simulated the circuit shown in Fig.~\ref{teleport}(a), assuming perfect state preparation of the left pair, but including all other sources of error. Based on our simulations, we expect that the spin in dot 4 will be in the $\ket{\U}$ state after the teleport with a probability of about 0.9, given a singlet on the left pair. In the presence of realistic hyperfine gradients (tens of MHz) and exchange strengths (several hundred MHz), we estimate that readout errors contribute the majority of the error.  

Assuming perfect preparation of a separated singlet state, our simulations suggest that the fidelity of the entanglement swap [Fig.~\ref{entangle}(a)] on a singlet state can be about 0.7, provided that the state is allowed to evolve in the presence of a quasi-static magnetic gradient to undo the coherent singlet-triplet evolution incurred during the SWAP operation in the presence of a gradient~\cite{Kandel2019} (see Supplementary Note 3). In this case, readout errors, state preparation errors of the EPR pair, and errors in the SWAP gate due to the magnetic gradient all contribute to the overall error. The average classical limit for teleporting entangled states of the type we use in this experiment is 2/3 ~\cite{Massar1995} (see Supplementary Note 2). By fitting the data of Fig.~\ref{entangle}(b) (see Methods, Supplementary Note 3, and Supplementary Fig.~9), we can also extract a maximum singlet teleportation probability of $0.71 \pm 0.04$, which compares favorably with the classical limit, although further research involving quantum state tomography is required to provide definitive proof. This value also agrees with our simulated fidelity and indicates that a classical explanation for our data is extremely unlikely (see Supplementary Notes 1 and 4). 

\section{Discussion}
This teleportation protocol is fully compatible with all gate-defined quantum-dot types, including Si quantum dots. Indeed, this teleportation protocol will work best with small magnetic gradients, as can be achieved with Si qubits. In large gradients,  resonant approaches~\cite{Nichol2017,Sigillito2019} or dynamically corrected gates~\cite{Wang2012} can still generate high-fidelity SWAP operations. State preparation errors can be suppressed by improving the coupling between the quantum dots and the reservoirs, and readout errors can be minimized by optimizing the position of the sensor quantum dots. We discuss the potential application of this technique to Si qubits in Methods.

As mentioned above and discussed further in Methods, the conditional quantum teleportation protocol we have developed is compatible with arbitrary qubit states. Deterministic quantum teleportation of arbitrary quantum states can also be realized with measurements of each qubit in the computational basis, together with CNOT~\cite{Zajac2018,Huang2019,Watson2018} and single-qubit gates~\cite{Yoneda2018}, which will enable complete measurements in the Bell-state basis~\cite{Nielsen2011}. Fast spin measurements together with real-time adaptive control~\cite{Shulman2014} could be used to complete the deterministic state transfer process.

The evidence we have presented for conditional state teleportation, entanglement swapping, and gate teleportation adds time-honored capabilities to the library of quantum information processing techniques available to spin qubits in quantum dots. Our results also highlight the potential of exchange-coupled spin chains for quantum information transfer. We envision that teleportation will be useful for the creation and manipulation of long-range entangled states and for error correction in quantum-dot spin qubits. As spin-based quantum information processors scale up, maintaining high-connectivity between spins will be critical, and quantum teleportation also opens an essential pathway toward achieving this goal. In many ways, spin qubits in quantum dots are an ideal platform for quantum teleportation, because they offer a straightforward means of generating and measuring entangled states of spins. As a result, we expect that quantum teleportation will find significant use in future spin-based quantum information processing efforts. 

\section{Methods}
\subsection{Device}
The four-qubit processor is a quadruple quantum dot fabricated on a GaAs/AlGaAs hetereostructure with a two-dimensional electron gas located 91 nm below the surface. The Si-doped region has vertical width of 14.3 nm, centered 24 nm below the top surface of the wafer. In this region, the dopant density is $3 \times 10^{18}$ cm$^{-3}$. The two-dimensional electron gas density $n=1.5 \times 10^{11}$cm$^{-2}$ and mobility $\mu=2.5 \times 10^6$cm$^2$V$^{-1}$s$^{-1}$ were measured at $T=4$K. 

Quantum dot fabrication proceeds as follows. Following ohmic contact fabrication via a standard metal stack and anneal, 10 nm of Al$_2$O$_3$ was deposited via atomic layer deposition. Three layers of overlapping aluminum gates~\cite{Angus2007,Zajac2015} were defined via electron beam lithography, thermal evaporation, and liftoff. The gate layers are isolated by a thin native oxide layer. The active area of the device is also covered with a grounded top gate. This is likely to screen the effects of disorder imposed by the oxide.  Empirically, we find that overlapping gates are essential for the exchange pulses we use in this work. The quadruple dot is cooled in a dilution refrigerator to a base temperature of approximately 10 mK. An external magnetic field $B=0.5$ T is applied in the plane of the semiconductor surface perpendicular to the axis connecting the quantum dots. Using virtual gates~\cite{Baart2016,Mills2019}, we tune the device to the single-occupancy regime. 

\subsection{Initialization}
To load the $\ket{T_{+,12}}=\ket{\U}_1 \ket{\U}_2$ state, we exchange electrons with the reservoirs in the (1,1) charge configuration~\cite{Foletti2009}. Both the magnetic field and temperature limit the fidelity of this process (Supplementary Fig.~10). We simulated the initialization fidelity by calculating the time-dependent populations of all relevant spin-states during the loading procedure. This simulation process is detailed in Ref.~\cite{Orona2018}. We assumed an electron temperature of 75 mK and a magnetic field of 0.5 T [Supplementary Fig.~10(a)]. This is broadly consistent with the electron temperatures we have measured in our setup, which range from 50-100 mK. Based on these simulations, we estimate that this state preparation fidelity is about $0.7$. The simulations presented here take this preparation error into account. In principle, increasing the magnetic field should improve the fidelity of the $\ket{T_+}$ loading process. Empirically, however, we did not observe a substantial enhancement with fields up to 1 T, as has previously been observed~\cite{Orona2018}. We suspect that unintentional dynamic nuclear polarization significantly modifies the magnetic field at the location of each dot. 

To load a separated singlet state, we exchange electrons with the reservoirs in the (0,2) charge configuration~\cite{Foletti2009}. We initialize the right pair of electrons in dot 4 as a singlet with $0.89$ probability for a load time of $2~\mu$s [Supplementary Fig.~10(b)]. This could be improved in the future by optimizing the coupling of the electrons to the source and drain reservoirs. Based on simulations of the Landau-Zener tunneling process to separate the electrons, we estimate that separating the singlet state incurs only a few percent error.

We can initialize either pair of electrons as $\ket{\D \U}$ or $\ket{\U \D}$ by adiabatically separating a singlet state~\cite{Foletti2009}. The orientation of the two spins in this product state depends on the orientation of the local hyperfine field.

\subsection{Exchange}
We induce exchange coupling between pairs of qubits by applying a voltage pulse to the barrier between the respective pair of dots~\cite{Reed2016,Martins2016}. Exchange coupling generated in this way is first-order insensitive to charge noise associated with the plunger gates. Barrier-gate pulses are accompanied by compensation pulses on the plunger gates to keep the dot chemical potentials fixed. For the exchange gates used in this work, we used a combination of barrier- ~\cite{Reed2016,Martins2016}, and tilt-controlled ~\cite{Petta2005} exchange. Empirically, we found that using this combination helps us to boost the exchange strength and improves the fidelity of the SWAP operation. All exchange pulses are optimized at the same tuning used to acquire all data in this work with one electron in each dot. We do not observe that pulsing exchange between two spins generates spurious enhanced exchange coupling elsewhere in the array.

\subsection{Readout}
Diabatic charge transfer into the outer dots projects the spin state of the separated pair onto the $\{\ket{S},\ket{T}\}$ basis~\cite{Petta2005,Foletti2009}. Adiabatic charge transfer into the outer dots maps either $\ket{\D \U}$ or $\ket{\U\D}$ to $\ket{S}$, depending on the sign of the local magnetic gradient, and it maps all other spin states to triplets~\cite{Petta2005,Foletti2009}. Here, ``diabatic'' or ``adiabatic'' refer to the speed with which the electrons are recombined relative to the size of the hyperfine gradient. We represent readout by diabatic charge transfer with an $``ST"$ in figures, and we represent readout by adiabatic charge transfer with a $``\D \U"$ in figures. When used to verify teleportation, diabatic charge transfer can only verify teleportation when $\ket{\phi}=\ket{\U}$. In principle, however, readout by adiabatic charge transfer could be used to measure qubit 4 in its computational basis. If $\Delta B_{34}$ were such that $\ket{\U}_3\ket{\D}_4$ were the ground state, adiabatic charge transfer would map $\ket{\U}_3\ket{\D}_4$ to a singlet, and $\ket{\U}_3\ket{\U}_4$ to a triplet. Together with tomographic rotation pulses, such a measurement would enable verification of teleportation of arbitrary states. 

In addition to conventional spin-blockade readout on both pairs of electrons, we use a shelving mechanism~\cite{Studenikin2012} to enhance the readout visibility. Using the two sensor quantum dots configured for rf-reflectometry (Fig.~\ref{apparatus})~\cite{Barthel2009}, we achieve single-shot readout with integration times of 4 $\mu$s on the left side and 6 $\mu$s on the right side and fidelities of 0.93 and 0.87, respectively. Relaxation times during readout were $65~\mu$s and $48~\mu$s on the left and right sides. Supplementary Figs.~12(a)-(b) show the experimentally measured curves demonstrating the relaxation during readout for both pairs of electrons. Supplementary Figs.~11(a)-(b) show fits to the readout histograms using Equations (1) and (2) in Ref.~\cite{Barthel2009} for each pair of qubits. In all teleportation measurements, both pairs of qubits are measured sequentially in the same single-shot sequence.

To determine the probabilities for the four different possibilities for joint measurements of both pairs, we fit the total measurement histogram for each pair separately. We determine the threshold for each pair by choosing the signal level that maximizes the visibility~\cite{Barthel2009}.  We then use these two thresholds to divide the probability distribution into quadrants. The overall probability is normalized, and we calculate the net probability in each quadrant. 

To eliminate any state-dependent crosstalk between qubit pairs during readout, we reload the first pair of electrons that we measure as an $\ket{S}$ before reading out the next pair for the data in Figs.~\ref{teleport} and \ref{exchange}. For the data in Fig.~\ref{entangle}, we additionally implemented a voltage ramp to bring each pair of electrons back to the (1,1) idling point immediately after readout. We empirically find that these procedures eliminate crosstalk during readout. The data in Supplementary Fig.~5 demonstrate that there is negligible readout or control crosstalk in our system.

Improvements to readout can be made by repositioning the sensor quantum dots for maximum differential charge sensitivity to achieve readout errors of $<0.01$ in integration times of $<1~\mu$s, as has previously been demonstrated in quantum dot spin qubits~\cite{Shulman2012,Shulman2014}.

\subsection{Simulation}
Our simulations include errors associated with state preparation, readout fidelity, relaxation during readout, charge noise, and the fluctuating magnetic gradient. We approximate singlet loading error by creating a two-electron state
\begin{equation}
\ket{\tilde{S}}=s_1\ket{S}+s_2\ket{T_0}+s_3\ket{T_+}+s_4\ket{T_-},
\end{equation}
where $|s_1|^2=f_\tn{s}$, and $|s_2|^2=|s_3|^2=|s_4|^2=(1-f_\tn{s})/3$. Also, $\ket{T_-}=\ket{\D \D}$, and $\ket{T_+}=\ket{\U \U}$. $f_\tn{s}=0.89$ is the singlet load fidelity. All coefficients are given random phases for each realization of the simulation. To simulate loading error during adiabatic separation of electrons, we set 
\begin{equation}
\ket{\tilde{G}}=s_1\ket{\D \U}+s_2\ket{\U \D}+s_3\ket{T_+}+s_4\ket{T_-},
\end{equation}
where the coefficients are the same as described above. We use the same coefficients, because the singlet initialization error dominates the error in this process. We also allow the orientation of the spins in this state to change between runs of the simulation as the hyperfine gradient changes. We approximate the $\ket{T_+}$ loading error by simulating the loading process as described in Ref.~\cite{Orona2018}. We directly extract the population coefficients of the other three two-electron spin states. We create a state which is a sum of all two-electron spin states:
\begin{equation}
\ket{\tilde{T_+}}=t_1\ket{S}+t_2\ket{T_0}+t_3\ket{T_+}+t_4\ket{T_-},
\end{equation}
where $|t_i|^2$ is determined as discussed above. We assign random phases to each of the coefficients during each realization of the simulation. 

To simulate the spin-eigenstate teleport operation, we set the initial state of the four qubit system as
\begin{equation}
\ket{\psi_i}=\ket{\tilde{T}_{+,12}}\otimes \ket{\tilde{S}_{34}}.
\end{equation}
To simulate the mixed-state and entangled-state teleport operations, we set the initial state as 
\begin{equation}
\ket{\psi_i}=\ket{\tilde{G}_{12}}\otimes \ket{\tilde{S}_{34}}.
\end{equation}

We incorporate charge noise and the hyperfine magnetic field and their effects on the SWAP operation by directly solving the 
Schr\"odinger equation for a four spin system. We generated a simulated SWAP operation from the following Hamiltonian: 
\begin{eqnarray}
	H_\tn{S}&=\frac{h}{4}J_{23} ( \sigma_{x,2} \otimes \sigma_{x,3}+\sigma_{y,2} \otimes \sigma_{y,3} + \sigma_{z,2} \otimes \sigma_{z,3})\\ &+\frac{g \mb}{2}\sum_{k=1}^4 B_k \sigma_{z,k} \label{Hamiltonian}
\end{eqnarray}
We assume a fixed exchange coupling of $J_{23}$ of 250 MHz between spins $2$ and $3$, and we adjust the time $T_S$ for the SWAP operation to give a $\pi$ pulse. These parameters correspond closely to the actual experiments. To account for charge noise, we allow the value of $J_{23}$ to fluctuate by 1$\%$ between simulation runs. We arrive at this level of charge noise via the expression $Q=\frac{J}{\sqrt{2}\pi \delta J}$~\cite{Dial2013}, where $\frac{\delta J}{J}$ is the fractional electrical noise, using the measured quality factor of 21. For the spin-eigenstate simulation, we set the local nuclear magnetic fields $B_k$ of spin $k$ to be $(-1,6,-4,0)$ MHz $\times \frac{2 h}{g \mb}$ for the qubits. We also include for each qubit the overall background field of 0.5 T. We allow the nuclear field at each site to fluctuate according to a normal distribution with standard deviation of 12 MHz for qubits 1 and 2 and 10 MHz for qubits 3 and 4. The field and fluctuations are adjusted to improve the agreement between the simulations in Supplementary Fig.~2 and Fig.~3. We empirically observe that the hyperfine fields fluctuate during the course of a given data-taking run, and they can even switch sign. Because we do not know a-priori what the hyperfine fields will be, it seems reasonable to treat them as fit parameters, especially since the chosen values fall well within the expected range. The overall evolution of the four-qubit system during the SWAP operation is given by the following propagator: $S_\tn{23}=\exp{\left(\frac{-i H_\tn{S} T_\tn{S}}{\hbar}\right)}$. 

The voltage pulses in our setup have finite rise times, which cause the four-qubit system to evolve under the magnetic gradient in the absence of exchange. To simulate this effect, we define 
\begin{eqnarray}
H_\tn{B}&=\frac{g \mb}{2}\sum_{k=1}^4 B_k \sigma_{z,k}.
\end{eqnarray}
Under this Hamiltonian, the wavefunction evolves according to the following propagator: $U_\tn{B}=\exp{\left(\frac{-i H_\tn{B} T_\tn{B}}{\hbar}\right)}$. In the experiment, all pulses are convolved in software with a Gaussian of width 2 ns before delivery to the qubits, so we set $T_\tn{B}=2$ ns. To simulate the spin-eigenstate teleport experiment, the simulated final state after the teleport operation is thus $\ket{\psi}=U_\tn{B} S_{23} U_\tn{B} \ket{\psi_i}$. 

For the simulations presented in Supplementary Fig.~1 and Supplementary Fig.~2, we also accounted for imperfections in our pulsing by allowing for the singlet-triplet state vector to rotate slightly during pulses which should ideally be perfectly diabatic. For example, suddenly separating a singlet is usually accompanied by some evolution toward the ground state of the hyperfine field, because the pulse is not perfectly sudden. We account for this by allowing the effective singlet-triplet state vector to rotate by 7 degrees toward the ground state of the hyperfine field during sudden separation of the singlet and by -7 degrees during readout via diabatic charge transfer. This rotation is implemented as a rotation about the $y$ axis in the effective $S-T_0$ subspace for each pair of qubits. The $y$ axis is defined by the usual $S-T_0$ Hamiltonian: $J\sz +\Delta B \sx$. The rotation angle of 7 degrees was chosen to match an additional control data set in which we adiabatically measured a singlet prepared via diabatic separation. Ideally, this measurement yields a singlet probability of 0.5. In practice, the measured singlet probability is slightly larger than this due to pulse errors, and 7 degrees was chosen to match the observed return probability. 

To compute the expected probabilities in Fig.~\ref{teleport}(d)-(e), we calculate all pairs of two-qubit correlators: $C_{\alpha,\beta}=\braket{\psi|\alpha \otimes \beta} \braket{\alpha \otimes \beta|\psi}$, where $\alpha$ (qubits 1 and 2) and $\beta$ (qubits 3 and 4) can be any of  $\{\ket{S},\ket{T_+},\ket{T_0},\ket{T_-}\}$. We calculate the probabilities in Fig.~\ref{teleport} as 
\begin{eqnarray}
&\pss=C_{\ket{S},\ket{S}},\\
&\ptt=\sum_{\substack{\alpha  \neq\ket{S}\\ \beta \neq\ket{S}}}C_{\alpha,\beta},\\
&\pst =\sum_{\beta\neq \ket{S}}C_{\ket{S},\beta},\\
&\pts =\sum_{\alpha\neq\ket{S}}C_{\alpha,\ket{S}}.
\end{eqnarray}

To simulate readout errors, we define the $g_\tn{L(R)} =1-\exp(-t_\tn{m}^\tn{L(R)}/T_1^\tn{L(R)})$ to be the probabilities that the triplet state on the left (right) side will relax to the singlet during readout. Here $t_\tn{m}^\tn{L(R)}$ is the measurement time, and $T_1^\tn{L(R)}$ is the relaxation time, as discussed above. We also set $r_\tn{L(R)}=1-f_\tn{L(R)}$ as the probability that singlet or triplet on the left (right) side will be misidentified due to noise. Here $f_\tn{L(R)}$ is the measurement fidelity due to random noise on the left (right) pair. The experimentally measured probabilities are 
\begin{widetext}
\begin{eqnarray}
\pss^{'}&=(1-\RL-\RR)\pss+\GL\pts+\GR\pst+\RL\pts+\RR\pst,\label{eq:ss}\\
\pst^{'}&=(1-\RL-\RR)\pst+\GL\ptt-\GR\pst+\RL\ptt+\RR\pss,\\
\pts^{'}&=(1-\RL-\RR)\pts-\GL\pts+\GR\ptt+\RL\pss+\RR\ptt,\\
\ptt^{'}&=(1-\RL-\RR)\ptt-\GL\ptt-\GR\ptt+\RL\pst+\RR\pts.
\end{eqnarray}
\end{widetext}
The displayed probabilities in Fig.~\ref{teleport}(d) are $\pss^{'}$, $\pst^{'}$, $\pts^{'}$, and $\ptt^{'}$.

To simulate the data shown in Fig.~\ref{exchange}, we generate variable exchange propagators $U_{12}$ and $U_{34}$ using Hamiltonians analogous to Eq.~\ref{Hamiltonian} for exchange between qubits 1-2 and qubits 3-4. Probabilities were calculated as described above. For example, to generate the simulations in Supplementary Fig.~3(a), the final state is computed as $\ket{\psi}=U_\B U_{34} U_\B S_{23} U_\B \ket{\psi_i}$. We compute all possible correlators $C_{\alpha,\beta}$, where $\alpha$ is any of $\{\ket{S},\ket{T_+},\ket{T_0},\ket{T_-}\}$, and $\beta$ is any of $\{\ket{\D\U},\ket{T_+},\ket{\U \D},\ket{T_-}\}$ and extract probabilities as discussed above. The simulated data are averaged over 1000 realizations of magnetic and electrical noise and random state errors. We note that the ground state configuration ($\ket{\D \U}$ or $\ket{\U \D}$) is allowed to change in the simulation if the gradient changes sign due to random noise. The results of these simulations are shown in Supplementary Fig.~3, which shows the operator sequences and initial states used to simulate the data.

To simulate the data in Fig.~\ref{entangle}, we compute the final state as $\ket{\psi}=S_{12}^{1/2} U_\B S_{23} U_\B S_{12}^{1/2} U_\B^\tn{R}(t) \ket{\psi_i}$. Here $U_\B^\tn{R}(t)$ indicates that the right-pair of qubits evolves for a variable time $t$ in their magnetic gradient $\Delta B_{34}$. We compute all possible correlators $C_{\alpha,\beta}$, where $\alpha$ is any of $\{\ket{\D\U},\ket{T_+},\ket{\U \D},\ket{T_-}\}$, and $\beta$ is any of $\{\ket{S},\ket{T_+},\ket{T_0},\ket{T_-}\}$. For this simulation, magnetic gradients were chosen to match the observed frequencies, and the width of the hyperfine distribution was reduced to mimic the effects of averaging for only a few seconds and to match the observed decay. For these data, exchange strengths were chosen to be 90 MHz.

To simulate the ideal results in the absence of noise in Fig.~\ref{teleport}, Supplementary Figs.~4  and 6, we eliminated all preparation and readout errors, all noise sources, and we eliminated the 2-ns evolution periods, which account for pulse rise times. We also eliminated the effect of magnetic gradients during the SWAP pulses.

\subsection{Estimation of $\Delta B$ Frequencies}
To extract the oscillation frequencies of the data in Fig.~4 and Supplementary Fig.~8, we zero-padded each line (corresponding to an average of up to 256 single-shot repetitions of each evolution time) by 256 points and took the absolute value of the fast Fourier transform of this averaged time series.  We then found the frequency giving the peak value. To reduce the effects of noise, we rejected all repetitions giving frequencies larger than 100 MHz. To generate the displayed frequency vs. repetition number traces, we smoothed the frequency vs. repetition series with a moving 10-point average.

\subsection{Applicability to Si spin qubits}
All of the necessary steps for conditional teleportation, including  barrier-controlled exchange~\cite{Reed2016}, and readout and initialization via Pauli spin-blockade~\cite{Jones2019}, have already been demonstrated in Si quantum dots. In general, we expect teleportation to work even better in Si, where magnetic gradients and noise can be reduced. One potential challenge is the requirement for spin blockade; small valley splittings in Si can easily lift spin-blockade~\cite{borselli2011}. However, this challenge is easily overcome by operating the quantum dots at larger occupation numbers where the singlet-triplet energy splitting is dominated by the orbital energy spacing~\cite{Higginbotham2014,West2019}. Another potential complication for Si qubits is the frequent use of micromagnets, which generate intense magnetic field gradients, for single-spin control. In particular, strong magnetic gradients make pure exchange rotations challenging. However, resonant exchange gates~\cite{Nichol2017,Sigillito2019} or dynamically corrected gates~\cite{Wang2012} can still generate high-fidelity SWAP operations in large magnetic gradients.

\subsection{Extension to deterministic teleportation of arbitrary states}
Deterministic teleportation of arbitrary states requires the ability to distinguish all four Bell states and the ability to generate arbitrary input states to the teleport. Achieving complete readout in the Bell-state basis is most easily achieved with single-qubit and CNOT gates together with single-qubit readout~\cite{Nielsen2011}. High-fidelity single-qubit~\cite{Yoneda2018} and CNOT gates ~\cite{Zajac2018,Huang2019,Watson2018} have already been demonstrated in Si. Single-spin readout can be achieved via Pauli spin-blockade measurements with a known ancilla spin~\cite{Zheng2019}, as discussed in the Readout section above, and SWAP operations. Alternatively, spin-selective tunneling~\cite{Elzerman2004} can be used. Fast spin measurements~\cite{Connors2020} together with real-time adaptive control~\cite{Shulman2014} could be used to complete the deterministic state transfer process.

\subsection{Data Availability}
The data that support the findings of this study are available from the corresponding author upon reasonable request.

\section{References}

\section{Acknowledgments}
We thank Lieven Vandersypen and Joseph Ciminelli for valuable discussions.  This work was sponsored the Defense Advanced Research Projects Agency under Grant No. D18AC00025; the Army Research Office under Grant Nos. W911NF-16-1-0260, W911NF-19-1-0167, and W911NF-18-1-0178; and the National Science Foundation under Grant DMR-1809343. The views and conclusions contained in this document are those of the authors and should not be interpreted as representing the official policies, either expressed or implied, of the Army Research Office or the U.S. Government. The U.S. Government is authorized to reproduce and distribute reprints for Government purposes notwithstanding any copyright notation herein.

\section{Author Contributions}
S.K.M., A.N.J., and J.M.N conceptualized the experiment. Y.P.K, H.Q, and J.M.N. conducted the investigation. S.F., G.C.G., and M.J.M. provided resources and conducted investigation. All authors participated in writing. J.M.N. supervised the effort.

\section{Additional information}
\subsection{Competing Interests}
The authors declare no competing interests.

\end{document}


\title{Supplementary Information -\\
	Conditional Teleportation of Quantum-Dot Spin States}

\author{Haifeng Qiao}
\thanks{These authors contributed equally.}

\author{Yadav P. Kandel}
\thanks{These authors contributed equally.}

\affiliation{Department of Physics and Astronomy, University of Rochester, Rochester, NY, 14627 USA}

\author{Sreenath K. Manikandan}
\affiliation{Department of Physics and Astronomy, University of Rochester, Rochester, NY, 14627 USA}

\author{Andrew N. Jordan}
\affiliation{Department of Physics and Astronomy, University of Rochester, Rochester, NY, 14627 USA}
\affiliation{Institute for Quantum Studies, Chapman University, Orange, CA 92866, USA}

\author{Saeed Fallahi}
\affiliation{Department of Physics and Astronomy, Purdue University, West Lafayette, IN, 47907 USA}
\affiliation{Birck Nanotechnology Center, Purdue University, West Lafayette, IN, 47907 USA}

\author{Geoffrey C. Gardner}
\affiliation{Birck Nanotechnology Center, Purdue University, West Lafayette, IN, 47907 USA}
\affiliation{School of Materials Engineering, Purdue University, West Lafayette, IN, 47907 USA}

\author{Michael J. Manfra}
\affiliation{Department of Physics and Astronomy, Purdue University, West Lafayette, IN, 47907 USA}
\affiliation{Birck Nanotechnology Center, Purdue University, West Lafayette, IN, 47907 USA}
\affiliation{School of Materials Engineering, Purdue University, West Lafayette, IN, 47907 USA}
\affiliation{School of Electrical and Computer Engineering, Purdue University, West Lafayette, IN, 47907 USA}

\author{John M. Nichol}
\email{john.nichol@ur.rochester.edu}

\affiliation{Department of Physics and Astronomy, University of Rochester, Rochester, NY, 14627 USA}


\pacs{}

\maketitle

\newpage

\renewcommand{\figurename}{Supplementary Figure}
\renewcommand{\thefigure}{\arabic{figure}}
\setcounter{figure}{0}    

\renewcommand{\tablename}{Supplementary Table}
\renewcommand{\thetable}{\arabic{table}}
\setcounter{table}{0}

\section{Supplementary Note 1. Entanglement swapping}
Here we derive the expected measurement dynamics for the entanglement swapping experiment. We label the quantum dots respectively $1,~2,~3,$ and $4$. The electrons in dots $1$ and $2$ are initialized in the following two qubit entangled state,
\begin{equation}
|\psi_{12}\rangle = \frac{1}{\sqrt{2}}\bigg(e^{i\phi}|\uparrow\rangle_{1}|\downarrow\rangle_{2}-e^{-i\phi}|\downarrow\rangle_{1}|\uparrow\rangle_{2}\bigg),
\end{equation}
where we take $\phi=\frac{-\pi}{4}$. The electrons in quantum dots $3$ and $4$ are also initialized in a two qubit entangled state,
\begin{equation}
|\psi_{34}\rangle = \frac{1}{\sqrt{2}}\bigg(e^{i\chi(t)}|\uparrow\rangle_{3}|\downarrow\rangle_{4}-e^{-i\chi(t)}|\downarrow\rangle_{3}|\uparrow\rangle_{4}\bigg),
\end{equation}
where $\chi(t)$ is the time dependent phase generated by the magnetic field gradient $\Delta B_{34}$.
The joint spin state of four qubits is,
\begin{widetext}
\begin{eqnarray}
|\psi_{0}\rangle &=& \frac{1}{\sqrt{2}}\bigg(e^{i\phi}|\uparrow\rangle_{1}|\downarrow\rangle_{2}-e^{-i\phi}|\downarrow\rangle_{1}|\uparrow\rangle_{2}\bigg)\otimes\frac{1}{\sqrt{2}}\bigg(e^{i\chi(t)}|\uparrow\rangle_{3}|\downarrow\rangle_{4}-e^{-i\chi(t)}|\downarrow\rangle_{3}|\uparrow\rangle_{4}\bigg)\nonumber\\
&=&\frac{1}{2}\bigg(e^{i[\phi+\chi(t)]}|\uparrow\rangle_{1}|\downarrow\rangle_{2}|\uparrow\rangle_{3}|\downarrow\rangle_{4}-e^{-i[\phi-\chi(t)]}|\downarrow\rangle_{1}|\uparrow\rangle_{2}|\uparrow\rangle_{3}|\downarrow\rangle_{4}\nonumber\\
&+&e^{-i[\phi+\chi(t)]}|\downarrow\rangle_{1}|\uparrow\rangle_{2}|\downarrow\rangle_{3}|\uparrow\rangle_{4}-e^{i[\phi-\chi(t)]}|\uparrow\rangle_{1}|\downarrow\rangle_{2}|\downarrow\rangle_{3}|\uparrow\rangle_{4}\bigg).
\end{eqnarray}
\end{widetext}
In principle, entanglement swapping via joint measurements can be demonstrated on this quantum state, by jointly measuring the qubits 2 and 3 (or 1 and 4), but for the ease of  implementing the joint measurement in our four qubit processor, we apply a unitary SWAP operation between the qubits 2 and 3 that swaps the quantum states of electrons between quantum dots $2$ and $3$, yielding the modified quantum state,
\begin{widetext}
\begin{eqnarray}
|\psi_{0}'\rangle &=& \frac{1}{2}\bigg(e^{i[\phi+\chi(t)]}|\uparrow\rangle_{1}|\uparrow\rangle_{2}|\downarrow\rangle_{3}|\downarrow\rangle_{4}-e^{-i[\phi-\chi(t)]}|\downarrow\rangle_{1}|\uparrow\rangle_{2}|\uparrow\rangle_{3}|\downarrow\rangle_{4}\nonumber\\
&+&e^{-i[\phi+\chi(t)]}|\downarrow\rangle_{1}|\downarrow\rangle_{2}|\uparrow\rangle_{3}|\uparrow\rangle_{4}-e^{i[\phi-\chi(t)]}|\uparrow\rangle_{1}|\downarrow\rangle_{2}|\downarrow\rangle_{3}|\uparrow\rangle_{4}\bigg).
\end{eqnarray}
\end{widetext}
The electron in quantum dot 1 is now entangled to the electron in quantum dot 3, and the electron in quantum dot 2 is entangled to the electron in quantum dot 4.  When we perform a joint measurement of qubits 3 and 4 in the $\{\ket{S},\ket{T}\}$ basis, and look at cases where the outcome is  a singlet, we find that the conditional state of electrons in quantum dots 1 and 2 is,
\begin{eqnarray}
|\Phi_{12}^{S_{34}}\rangle=\langle S_{34}|\psi_{0}'\rangle &=& \frac{1}{2\sqrt{2}}\bigg(e^{i[\phi-\chi(t)]}|\uparrow\rangle_{1}|\downarrow\rangle_{2}-e^{-i[\phi-\chi(t)]}|\downarrow\rangle_{1}|\uparrow\rangle_{2}\bigg).
\end{eqnarray}
Recall $|S\rangle=\frac{1}{\sqrt{2}}\big(|\uparrow\downarrow\rangle-|\downarrow\uparrow\rangle\big)$. We find that, as a result of the joint measurement of the electrons in dots 3 and 4, in cases where the measurement outcome is a singlet entangled state, the reduced state of electrons in dots 1 and 2 is maximally entangled. Also note that the coherent singlet-triplet evolution previously occurring between the qubits 3 and 4 now happens between qubits 1 and 2, provided a singlet is measured on qubits 3 and 4. 

We may also derive the unconditional quantum density matrix for qubits  1 and 2, after an ideal joint measurement of qubits 3 and 4 in the $\{S,T\}$ basis. We obtain the following conditional states for qubits 1 and 2 for any of the triplet outcomes on qubits 3 and 4:
\begin{eqnarray}
|\Phi_{12}^{T^{0}_{34}}\rangle= \langle T^{0}_{34}|\psi_{0}'\rangle &=&-\frac{1}{2\sqrt{2}}\bigg(e^{i[\phi-\chi(t)]}|\uparrow\rangle_{1}|\downarrow\rangle_{2}+e^{-i[\phi-\chi(t)]}|\downarrow\rangle_{1}|\uparrow\rangle_{2}\bigg),\nonumber\\
|\Phi_{12}^{T^{+}_{34}}\rangle=  \langle T^{+}_{34}|\psi_{0}'\rangle &=& \frac{e^{-i[\phi+\chi(t)]}}{2}|\downarrow\rangle_{1}|\downarrow\rangle_{2}=\frac{e^{-i[\phi+\chi(t)]}}{2}\ket{T^-_{12}}\nonumber\\
|\Phi_{12}^{T^{-}_{34}}\rangle =  \langle T^{-}_{34}|\psi_{0}'\rangle &=& \frac{e^{i[\phi+\chi(t)]}}{2}|\uparrow\rangle_{1}|\uparrow\rangle_{2}=\frac{e^{i[\phi+\chi(t)]}}{2}\ket{T^+_{12}}.
\end{eqnarray}
We have denoted $\ket{T^0}=\frac{1}{\sqrt{2}}(\ket{\U \D}+\ket{\D \U})$, $\ket{T^+}=\ket{\U \U}$ and $\ket{T^-}=\ket{\D \D}$. To simplify notation, in this section we have distinguished the triplet states with superscripts.

It is easily verified that 
\begin{eqnarray}
|\Phi_{12}^{S_{34}}\rangle\langle \Phi_{12}^{S_{34}}|+|\Phi_{12}^{T^{0}_{34}}\rangle\langle \Phi_{12}^{T^{0}_{34}}|&=& \frac{1}{4} \bigg( \ket{\U \D}\bra{\U \D} +\ket{\D \U}\bra{\D \U}\bigg) \nonumber \\
&=& \frac{1}{8} \bigg( \left(\ket{S}+\ket{T^0}\right)\left(\bra{S}+\bra{T^0}\right) +\left(\ket{S}-\ket{T^0}\right)\left(\bra{S}-\bra{T^0}\right)  \bigg) \nonumber \\
&=&\frac{1}{4}  \left(\ket{S}\bra{S}+\ket{T^0}\bra{T^0}\right).  
\end{eqnarray}
The unconditioned state after measurement in the $\{S,T\}$ basis is therefore,
\begin{equation}
\hat{\rho}_{12}=|\Phi_{12}^{S_{34}}\rangle\langle \Phi_{12}^{S_{34}}|+|\Phi_{12}^{T^{0}_{34}}\rangle\langle \Phi_{12}^{T^{0}_{34}}|+|\Phi_{12}^{T^{+}_{34}}\rangle\langle \Phi_{12}^{T^{+}_{34}}|+|\Phi_{12}^{T^{-}_{34}}\rangle\langle \Phi_{12}^{T^{-}_{34}}|=\frac{\hat{\mathbb{I}}_{4\times 4}}{4}\end{equation}
in the $\{S,T\}$ basis. This density matrix represents the completely mixed state, having no time-dependence. 

\section{Supplementary Note 2. Classical Fidelity Bound for Entanglement Swapping}
In this section, we argue that the classical bound on the teleportation fidelity for the entanglement swap experiment is 2/3. 

The fidelity of classical teleportation of a single qubit is 2/3~\cite{Massar1995}. The reason for this is the following. If Alice wishes to teleport an unknown single qubit quantum state to Bob, she could simply measure her qubit (to the best of her abilities using generalized measurements) and transmit her result to Bob, who could then prepare his qubit according to Alice’s measurement. The average probability that Bob’s state matches Alice’s original state is 2/3. This fidelity estimate is based on the arguments of Ref.~\cite{Massar1995}, and it is the best one can achieve in the absence of systematic errors and not using additional quantum resources. A fidelity exceeding this bound, 2/3 for a single qubit, is a strong indication that additional quantum features are present in the experiment and has been used as a good benchmark to verify successful quantum teleportation.

To extend this result to entangled states of the type we use in this experiment, we note that each pair of qubits initially occupies the $m_s=0$ subspace, where $m_s$ is the total $z$-component of angular momentum. Therefore, the state of each pair of qubits is spanned by the $\ket{S}$ and $\ket{T_0}$ joint spin states and is a singlet-triplet qubit.  Assuming conservation of $m_s$, upon post selection of one pair as a singlet, the other pair thus must have $m_s=0$ and is also a singlet-triplet qubit. Thus, we can view teleportation of entangled states in our experiment as a transfer of singlet-triplet qubit states, and we may ask, how well can Alice classically transmit an individual singlet-triplet qubit. Note also that our experiment involves teleportation not of arbitrary singlet-triplet states, but only states of the form $\cos(\theta/2)\ket{S}+\exp(\pm i \pi/2)\sin(\theta/2)\ket{T_0}$.  Because the fidelity with which Bob can approximate Alice’s states does not depend on the azimuthal angle (here $\pm \pi/2$) of her state on her Bloch sphere, we expect that the relevant classical bound for this experiment is identical to the usual single-qubit result of 2/3. Below we provide a proof of this result.

Let $\ket{\psi}=\cos{(\theta/2)}\ket{0}+e^{i \phi} \sin{(\theta/2)} \ket{1}$ be an arbitrary qubit state. Suppose Alice measures $\ket{\psi}$, and finds $\ket{0}$. This occurs with probability $P_0=\cos^2{(\theta/2)}$. Bob then prepares his qubit as $\ket{0}$. The fidelity of Bob's qubit with respect to the initial state $\ket{\psi}$ is $S_0=|\braket{0|\psi}|^2=\cos^2{(\theta/2)}$. 

Suppose instead that Alice measures $\ket{\psi}$ and finds $\ket{1}$. This occurs with probability $P_1=\sin^2(\theta/2)$. In this case, the fidelity  of Bob's qubit with respect to the initial state is $S_1=|\braket{1|\psi}|^2=\sin^2(\theta/2)$. 

Define $S(\theta,\phi)=P_0S_0+P_1S_1=\sin^4(\theta/2)+\cos^4(\theta/2)$ to be the fidelity of Bob's qubit for a particular choice of $\theta$ and $\phi$. Thus, the average fidelity of Bob's qubit with respect to $\ket{\psi}$, averaged over all possible states, is 
\begin{eqnarray}
F&=&\frac{\int_{0}^{2 \pi} d\phi\int_{0}^{\pi} d \theta \sin(\theta) S(\theta,\phi)}{\int_{0}^{2 \pi} d\phi\int_{0}^{\pi} d \theta \sin(\theta)} = \frac{2}{3}.
\end{eqnarray}

We now proceed to show that the same bound holds for states $\ket{\psi}$ in the experiment in the $y-z$ plane containing the two poles of the Bloch sphere, $|0\rangle$ and $|1\rangle$. The fidelity computed for this sub-ensemble of states becomes,
\begin{equation}
	F'=\frac{\int_{0}^{2 \pi} d\phi\int_{0}^{\pi} d \theta \sin(\theta) S(\theta,\phi))[\delta(\phi-\pi/2)+\delta(\phi-3\pi/2)]}{ \int_{0}^{2 \pi} d\phi\int_{0}^{\pi} d \theta \sin(\theta)[\delta(\phi-\pi/2)+\delta(\phi-3\pi/2)]} =\frac{2}{3}.
\end{equation}
The above equation indicates that 2/3 is a classical bound for the  fidelity with respect to states in the $y-z$ plane.

Any realistic teleportation experiment involves choosing a subset of all possible input states.  A common choice is a complete set of orthogonal input states~\cite{Bouwmeester1997,Riebe2004,Olmschenk2009,Steffen2013,Pfaff2014,Pirandola2015, Reindl2018}.  As we have discussed in the paper, the input states available to us in this work correspond to singlet-triplet qubit states on the $y-z$ plane of the Bloch sphere. When choosing a subset of input states, one must take care not to assume any additional information about the teleportation protocol that could be advantageously used during verification. In the case of $F'$, for example, we do not assume that states following teleportation lie along a particular plane of the Bloch sphere. Full quantum state tomography is required to establish teleportation fidelity above the classical bound. Our assumption that teleportation can take qubit states out of the initial plane justifies the spherical integral discussed above.

\section{Supplementary Note 3. Estimated Fidelity}
We simulate the fidelity of the conditional teleport operation for spin eigenstates by starting with the initial state $\ket{\psi_i}=\ket{T_{+,12}}\otimes \ket{\tilde{S}_{34}}$, and we simulate the following operator sequence, as discussed above: $\ket{\psi}=U_B S_{23} U_B \ket{\psi_i}$. We seek to compute the probability that a singlet outcome on the left pair coincides with qubit 4 having the $\ket{\U}$ state. To assess this probability, we compute all correlators $C_{\alpha,\beta}$ as before. To incorporate readout errors and relaxation, we set
\begin{equation}
C_{\ket{S},\beta}^{'}=(1-r_L)C_{\ket{S},\beta}+(g_L+r_L)\sum_{\alpha\neq\ket{S},\beta } C_{\alpha,\beta}.
\end{equation}

The probability for qubit 4 to have the $\ket{\U}$ state, conditioned on the left pair yielding a singlet, is
\begin{equation}
P_{\U}=\frac{C_{\ket{S},\ket{T_+}}^{'}+\frac{1}{2}\left(C_{\ket{S},\ket{S}}^{'}+C_{\ket{S},\ket{T_0}}^{'}\right)}{ \sum_{\beta } C_{\ket{S},\beta}^{'}}.
\end{equation}
We averaged $P_{\U}$ over 100 different realizations of the magnetic and electric noise and state preparation errors, as discussed above.

The fidelity of the combined teleport and SWAP operation for entangled states is determined through a simulation similar to that described above. We initialized the array in the $\ket{\psi_i}=\ket{\tilde{G}_{12}}\otimes \ket{S_{34}}$ state, and the final state is computed as $\ket{\psi}=S_{12}^{1/2}U_B S_{23} U_B S_{12}^{1/2} U_B^R(t) \ket{\psi_i}$. We have assumed perfect state preparation on the right pair of qubits. Now, we compute all possible correlators $C_{\alpha,\beta}$, where $\alpha$ and $\beta$ are any one of $\{\ket{S},\ket{T_+},\ket{T_0},\ket{T_-}\}$. We incorporate readout errors as above, and the fidelity was computed as the maximum value of $C_{\ket{S},\ket{S}}^{'}$ as the evolution time $t$ varies in a quasi-static gradient.  Choosing the maximum value in this way is equivalent to picking a single-qubit $z$ rotation to undo the effects of singlet-triplet evolution in a gradient.

A first estimate of the fidelity for singlet teleportation may be obtained from our entanglement swap data $p(S_L|S_R)$ by determining the maximum value of $p(S_L|S_R)$ while correcting for left-side readout errors and right-side state preparation errors. To correct the $p(S_L|S_R)$ data for left side readout errors, we invert Eq.~\ref{eq:ss} to obtain
\begin{eqnarray}
P_{SS}=\frac{P_{SS}'-g_L-r_L}{1-g_L-2r_L}. \label{ss2}
\end{eqnarray}
In writing this equation, we have set $r_R=g_R=0$ and $P_{TS}=1-P_{SS}$, since we neglect right-side readout errors and only consider those experimental runs yielding a singlet on the right side. We apply Eq.~\ref{ss2} to our data ($P_{SS}'$) to extract $P_{SS}$. We then fit the first period of the resulting data to a function of the form $P(t)=A+B\cos(\omega t +\delta)$, where $A$, $B$, $\omega$, and $\delta$ are fit parameters. To correct for preparation errors of the state to be teleported, we set $B'=B/(2f_s-1)$, where $f_s$ is the singlet load fidelity, discussed above. We compute the maximum singlet return probability as $P(15~ \textnormal{ns})= 0.71 \pm 0.04$ with $B'$ used in place of $B$. The quoted uncertainty is the $95\%$-confidence interval associated with the fit. The data and fits are shown in Supplementary Fig.~\ref{swapFit}.

\section{Supplementary Note 4. Hypothesis testing}
We consider the following two hypotheses for the time-series data presented as verification of entanglement swapping,
\begin{itemize}
	\item Hypothesis Q: Entanglement swapping is achieved using quantum correlations as a resource. This means that the measurement outcome on the left pair of qubits depends on the measurement outcome of the right pair of qubits in a predictable way.
	\item Hypothesis C: The quantum information is transmitted via classical means. This means that the measurement outcomes on the left pair of qubits would not have a conditional dependence on the measurement outcome on the right pair of qubits. 
\end{itemize}

The key test of quantum effects in this experiment is thus the observation of oscillations on the left pair of qubits that appear only when the left-side measurements are conditioned on the right-side measurements. On the one hand, when entanglement swapping is successful, $p(S_L|S_R)$ and $p(S_L|T_R)$ (shown in Fig. 4(b) in the main text and in Supplementary Fig.~\ref{pval} here) should both oscillate in time as the input state $\ket{\psi}$ precesses in the $y-z$ plane of the $S-T_{0}$- qubit Bloch sphere. This oscillation occurs even though the unconditioned data $p(S_L)$ show no oscillations. The absence of oscillations in the unconditioned data happens because on average, Bob's member of the EPR pair and the single-qubit state to be teleported (a member of another entangled pair) are completely uncorrelated, as discussed in detail above.

On the other hand, if the unconditioned data $p(S_L)$  were to show the same oscillation as $p(S_L|S_R)$, for example, the claim of conditional entanglement swapping using quantum resources would be invalidated. This could occur, for instance, when the oscillations on the left pair of qubits are a result of a local preparation/driving when the information about the quantum state on the right pair is received via classical means (i.e. classical teleportation). 

As discussed in the caption to Supplementary Fig. 6, the amplitude of oscillations associated with $p(S_L|T_R)$ is expected to be smaller than the oscillations associated with $p(S_L|S_R)$ because our measurements do not distinguish the three triplets, which are linear combinations of the three Bell states $\ket{\Psi^+}$,  $\ket{\Phi^+}$, and $\ket{\Phi^-}$. 

To assess the probability of the classical hypothesis (C), we first quantify the amplitude of the residual oscillations in the unconditioned data $p(S_L)$, which may occur because of incomplete averaging, or an imperfect SWAP.  We fit these data to a function of the form $g(t)=p_0+A e^{-t/\tau}\cos(2 \pi f t+\phi)$, where $p_0$, $A$, $\tau$, $\omega$, and $\phi$ are fit parameters. The unconditioned data have an amplitude $A_U=0.030 \pm 0.007$. Here the uncertainty is a standard error $\sigma_U$. Similarly, to test the quantum hypothesis (Q), we fit the data $p(S_L|S_R)$ and $p(S_L|T_R)$ to the same function, and we extract  $A_S=0.201 \pm 0.021$ and $A_T=0.102 \pm 0.009$. The uncertainties are the standard errors $\sigma_S$ and $\sigma_T$, respectively. Our fitting routine in Matlab computes the confidence intervals using the $t$-distribution. Using the $t$-distribution, we convert the confidence interval to a standard error. Table~\ref{tab1} shows the fitted parameters, standard errors $\sigma$, and  $95\%$ confidence intervals $c$. Note the close agreement between the usual relation $c=1.96 \sigma$ for a normal distribution.  Note that $\sigma_U<\sigma_S \ll A_S-A_U$, and $\sigma_U<\sigma_T \ll A_T-A_U$, suggesting a very low probability for the classical hypothesis.

Define
\begin{eqnarray}
G(x,\sigma,\mu)=\frac{1}{\sqrt{2 \pi} \sigma} e^{-\frac{1}{2}\left(\frac{x-\mu}{\sigma}\right)^2}.
\end{eqnarray}
The probability of the classical hypothesis for the case of singlet conditioning is the probability that the amplitude of $p(S_L|S_R)$ is less than $A_U$. Assuming that the amplitude values of $p(S_L|S_R)$ are normally distributed between experimental runs, we assess this probability as 
\begin{eqnarray}
p_S=\int_{-\infty}^{ A_U} G\left(x,\sigma_S,A_S\right) dx < 0.00001.
\end{eqnarray}
Note that since $\sigma_S > \sigma_U$, it is reasonable to neglect the variance of $A_U$ to a first approximation. The probability of a classical explanation for the case of triplet conditioning is the probability that the amplitude of $p(S_L|T_R)$ is less than $A_U$. We assess this probability as 
\begin{eqnarray}
p_T=\int_{-\infty}^{ A_U} G\left(x,\sigma_T,A_T\right) dx < 0.00001.
\end{eqnarray}
Since $\sigma_T < \sigma_U$, it is also reasonable to neglect the variance of $A_C$ to a first approximation. 

To account for the variance of $A_U$, we may perform Welch's $t$-test accounting for unequal sample variances,
\begin{eqnarray}
t_S= \frac{A_S-A_U}{\sqrt{\sigma_U^2+\sigma_S^2}}= 7.72,
\end{eqnarray}
and
\begin{eqnarray}
t_T= \frac{A_T-A_U}{\sqrt{\sigma_U^2+\sigma_T^2}}= 6.32,
\end{eqnarray}
We use the Welch–Satterthwaite equation to estimate the number of degrees of freedom, finding 150 and 232 for the cases of singlet and triplet conditioning respectively. The $p$-values associated with the $t_S$ and $t_T$ scores both satisfy $p<0.00001$. Given these probabilities, a classical explanation for the data is extremely unlikely. 

A further, more conservative estimate for the classical hypothesis probability is obtained by analyzing all of the data shown in Supplementary Fig. 7. For each repetition $i=1,2,\cdots,256$, we calculate $U_i=\frac{\max(p(S_L))-\min(p(S_L))}{2}$, $S_i=\frac{\max(p(S_L|S_R))-\min(p(S_L|S_R))}{2}$, and $T_i=\frac{\max(p(S_L|T_R))-\min(p(S_L|T_R))}{2}$. $S_i$, $T_i$, and $U_i$ quantify the maximum amplitude of the singlet-conditioned, triplet-conditioned, and unconditioned time series data for each repetition $i$.

Supplementary Figure~\ref{pval2} shows the distributions of $S_i$, $T_i$, and $U_i$ over all 256 different repetitions. We calculate $\mu_S=0.234$, $\sigma_S=0.036$, $\mu_T=0.162$, $\sigma_T=0.021$, $\mu_U=0.113$, and $\sigma_U=0.015$. Here, $\mu$ indicates an average, and $\sigma$ refers to the estimated standard deviation.  As before, we compute the $t$-scores associated with these distributions:
\begin{eqnarray}
t_S= \frac{\mu_S-\mu_U}{\sqrt{\sigma_U^2/N+\sigma_S^2/N}}= 49.46,
\end{eqnarray}
and
\begin{eqnarray}
t_T= \frac{\mu_T-\mu_U}{\sqrt{\sigma_U^2/N+\sigma_T^2/N}}= 29.92,
\end{eqnarray} 
where $N=256$ is the sample size. These large $t$-score values result from the large sample sizes and well-separated distributions. Using the Welch–Satterthwaite equation, we find the number of degrees of freedom to be 339  and 463 for singlet and triplet conditioning. The $p$-values associated with these $t$-scores both satisfy $p<0.00001$. Both $p$-values are below 0.05, indicating that the classical hypothesis is extremely unlikely.

We may assess the impact of potential spurious classical correlations in our data as follows. As discussed in the main text, the data shown in Supplementary Fig.~\ref{swapcontrol} represent the same experiment as the entanglement swap, but the SWAP gate to distribute the entangled pair was omitted. Thus, the left-side measurements in Supplementary Fig.~\ref{swapcontrol}(a) provide a bound on the presence of oscillations caused by classical means, including readout cross-talk. We have fit these data as described above in this section, and as shown in Supplementary Fig.~\ref{pvalU}. The fitted parameters are shown in Supplementary Table~\ref{tab2}.

We may determine if the observed oscillations in the control experiment are statistically different from the oscillations associated with the full entanglement-swap data using the following $t$ scores:
\begin{eqnarray}
t_S^c= \frac{A_S-A_S^c}{\sqrt{\sigma_S^2+(\sigma_S^c)^2}}= 7.01,
\end{eqnarray}
and
\begin{eqnarray}
t_T^c= \frac{A_T-A_T^c}{\sqrt{\sigma_T^2+(\sigma_T^c)^2}}= 3.05,
\end{eqnarray}
Here, the variables with a ``c'' superscript are obtained from Supplementary Table~\ref{tab2}. In both of these cases, we obtain the degrees of freedom as discussed above, and the $p$-values associated with the $t$ scores both satisfy $p<0.005$. Given these probabilities, it is unlikely that classical correlations could explain our data.

\newpage

\begin{table*}
	\centering
	\begin{tabular}{|c|c|c|c|c|c|c|c|c|c|}
		\hline
		&\multicolumn{3}{c|}{$p(S_L)$}&\multicolumn{3}{c|}{$p(S_L|S_R)$} &\multicolumn{3}{c|}{$p(S_L|T_R)$} \\ 
		\hline
		Parameter & value & $c$& $\sigma$&value&$c$ &$\sigma$&value&$c$ &$\sigma$   \\ 
		\hline
		$p_0$       & 0.362 & 0.007& 0.004& 0.464 &0.011 &0.006 & 0.287 &0.008 &0.004\\ \hline
		$A$         & 0.030 &0.014 & 0.007& 0.201 &0.042 &0.021 & 0.102 &0.017 &0.009\\ \hline 
		$f$ (GHz)   & 0.026 & .002 & 0.001& 0.028 &0.0005&0.0003& 0.029 &0.0008&0.0004 \\ \hline
		$\phi$      & 4.54  & 0.76 & 0.38 & 3.84  &0.19  &0.10   & 0.557 &0.271&0.137\\ \hline
		$\tau$ (ns) & 151   &0     &0     & 142   & 71   &36    & 151   & 0    &0\\ 
		\hline
		
	\end{tabular}
	\caption{Fitted parameters. $c$ denotes half of the $95\%$-confidence interval width, and $\sigma$ is the standard error. In the case of the unconditioned data $p(S_L)$ and $p(S_L|T_R)$, we did not fit for the decay time $\tau$. Instead, we fixed it at the value obtained from the fit to $p(S_L|S_R)$.}
	\label{tab1}
\end{table*}

\begin{table*}
	\centering
	\begin{tabular}{|c|c|c|c|c|c|c|c|c|c|}
		\hline
		&\multicolumn{3}{c|}{$p(S_L)$}&\multicolumn{3}{c|}{$p(S_L|S_R)$} &\multicolumn{3}{c|}{$p(S_L|T_R)$} \\ 
		\hline
		Parameter & value & $c$& $\sigma$&value&$c$ &$\sigma$&value&$c$ &$\sigma$   \\ 
		\hline
		$p_0$       & 0.534 & 0.006& 0.003& 0.538 &0.008 &0.004 & 0.521 &0.010 &0.005\\ \hline
		$A$         & 0.010 &0.015 & 0.008& 0.038 &0.019 &0.010 & 0.056 &0.024 &0.012\\ \hline 
		$f$ (GHz)   & 0.027 & 0 & 0& 0.027 &0     &0     & 0.027 & 0    &0 \\ \hline
		$\phi$      & 4.46  & 1.51 & 0.763 & -0.92  &0.54  &0.27   & 2.38 &0.47 &0.24\\ \hline
		$\tau$ (ns) & 78   &0     &0     & 78   & 0   &0    & 78   & 0    &0\\ 
		\hline
		
	\end{tabular}
	\caption{Fitted parameters for the control experiment. $c$ denotes half of the $95\%$-confidence interval width, and $\sigma$ is the standard error. We used the data in Supplementary Fig.~\ref{swapcontrol}(b) to fix the values of $f$ and $\tau$ in the fits. }
	\label{tab2}
\end{table*}

\newpage

\begin{figure*}
	\centering
	\includegraphics[width=6.5in]{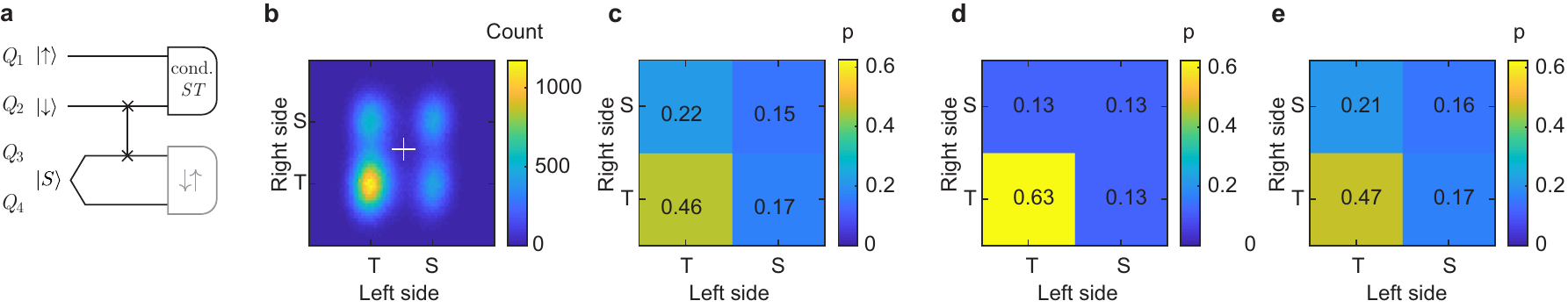}
	\caption{\label{mixed}  Conditional teleportation of a classical mixed spin state. (a) Quantum circuit used. The conditional singlet-triplet measurement on qubits 1 and 2 induces teleportation, and the gray measurement of the right pair verifies teleportation. In contrast to the experiment of Fig.~\ref{teleport}, in this experiment, we adiabatically separate the electrons in the left pair. This process generates either $\ket{\D}_1 \ket{\U}_2$ or $\ket{\U}_1 \ket{\D}_2$, depending on the sign of the hyperfine gradient $\Delta B_{12}$. The hyperfine gradient fluctuates randomly during the course of the experiment, leading to the generation of a classical mixed state on qubit 1 (and qubit 2). Likewise, the measurement on the right pair projects either $\ket{\D}_3 \ket{\U}_4$ or $\ket{\U}_3 \ket{\D}_4$ to $\ket{S_{34}}$, depending on the sign of $\Delta B_{34}$. (b) Experimentally measured probability distribution for $65{,}536$ single-shot realizations of the teleportation sequence in (a). The white cross indicates the threshold used to calculate probabilities. (c) Extracted probabilities $p$ from the distribution in (b). (d) Simulated probabilities computed neglecting any errors. (e) Simulated probabilities accounting for readout errors, state preparation errors, charge noise, and hyperfine fields. All probabilities are rounded to the nearest hundredth.}
\end{figure*}

\begin{figure*}
	\centering
	\includegraphics{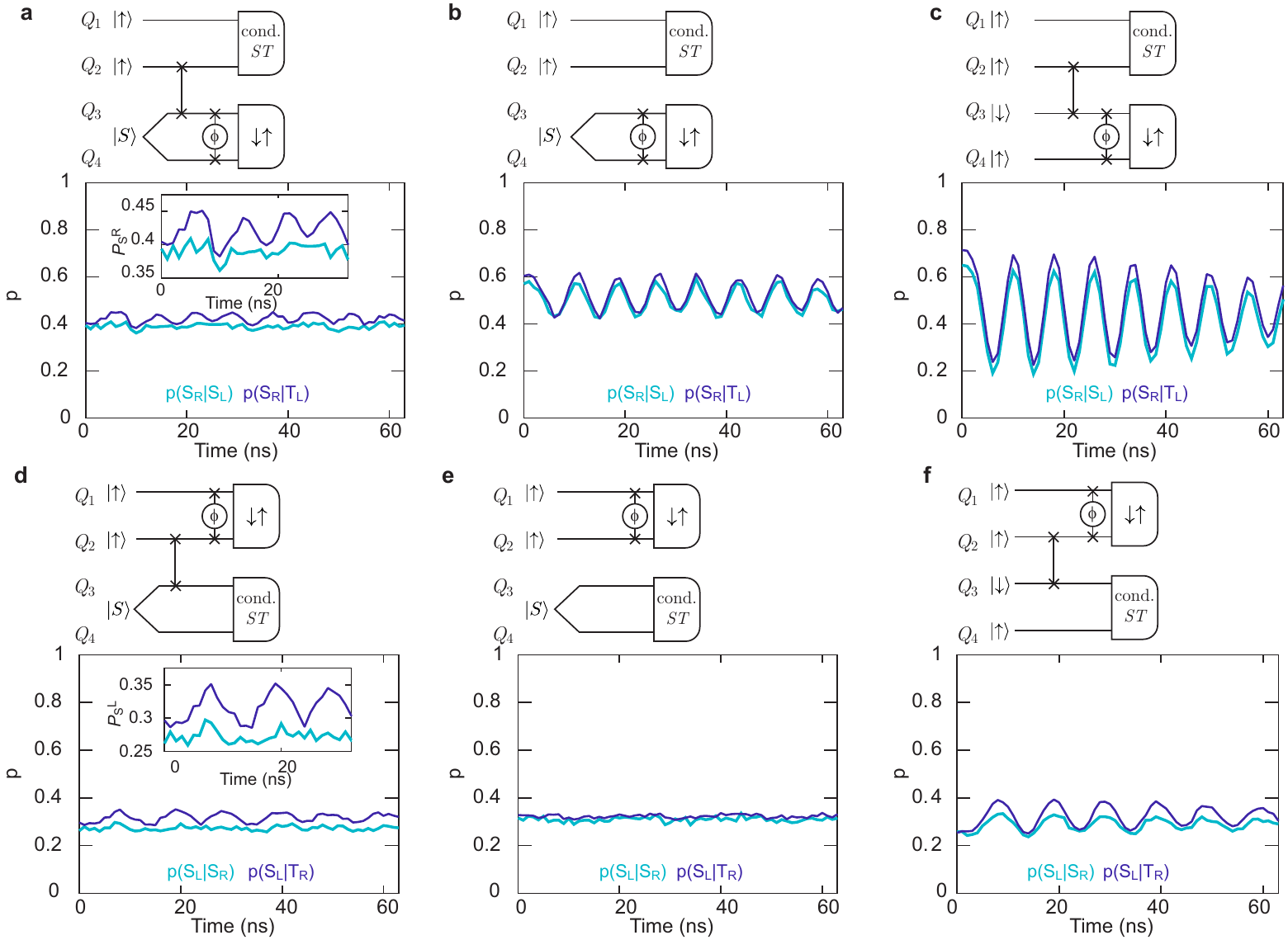}
	\caption{\label{fullExchange} Verification of conditional teleportation of a classical spin state. (a) We apply a variable exchange gate to the right pair after measuring the left pair. Here, $\phi=2\pi J_{34}t$, where $J_{34}$ is the exchange coupling between qubits 3 and 4, and $t$ is the evolution time given by the $x$-coordinate of each data point. ($\phi=\pi$ corresponds to a SWAP operation.)  When then left pair give a singlet, the right pair have the same spin, and no oscillations should be visible. The inset shows the same data from 0-32 ns. (b) Control experiment with no SWAP operation. Residual oscillations result from errors in preparing the EPR pair. (c) Control experiment with the EPR pair replaced by a product state. Oscillations occur because the fluctuating hyperfine gradient between the right pair sometimes favors the $\ket{\U \D}$ orientation. (d) Applying an exchange gate to the left pair after measuring the right pair generates exchange oscillations on the left pair only if the right pair yields a triplet. The inset shows the same data from 0-32 ns. (e) Control experiment with no SWAP operation. No oscillations occur because qubits 1 and 2 are prepared in the $\ket{\U \U}$ state. (f) Control experiment with the EPR pair replaced by a product state. Here, the oscillations are small because the hyperfine gradient between qubits 3 and 4 fluctuates around zero. Each data point represents the average of $16{,}384$ single shot measurements.}
\end{figure*}

\begin{figure*}
	\centering
	\includegraphics{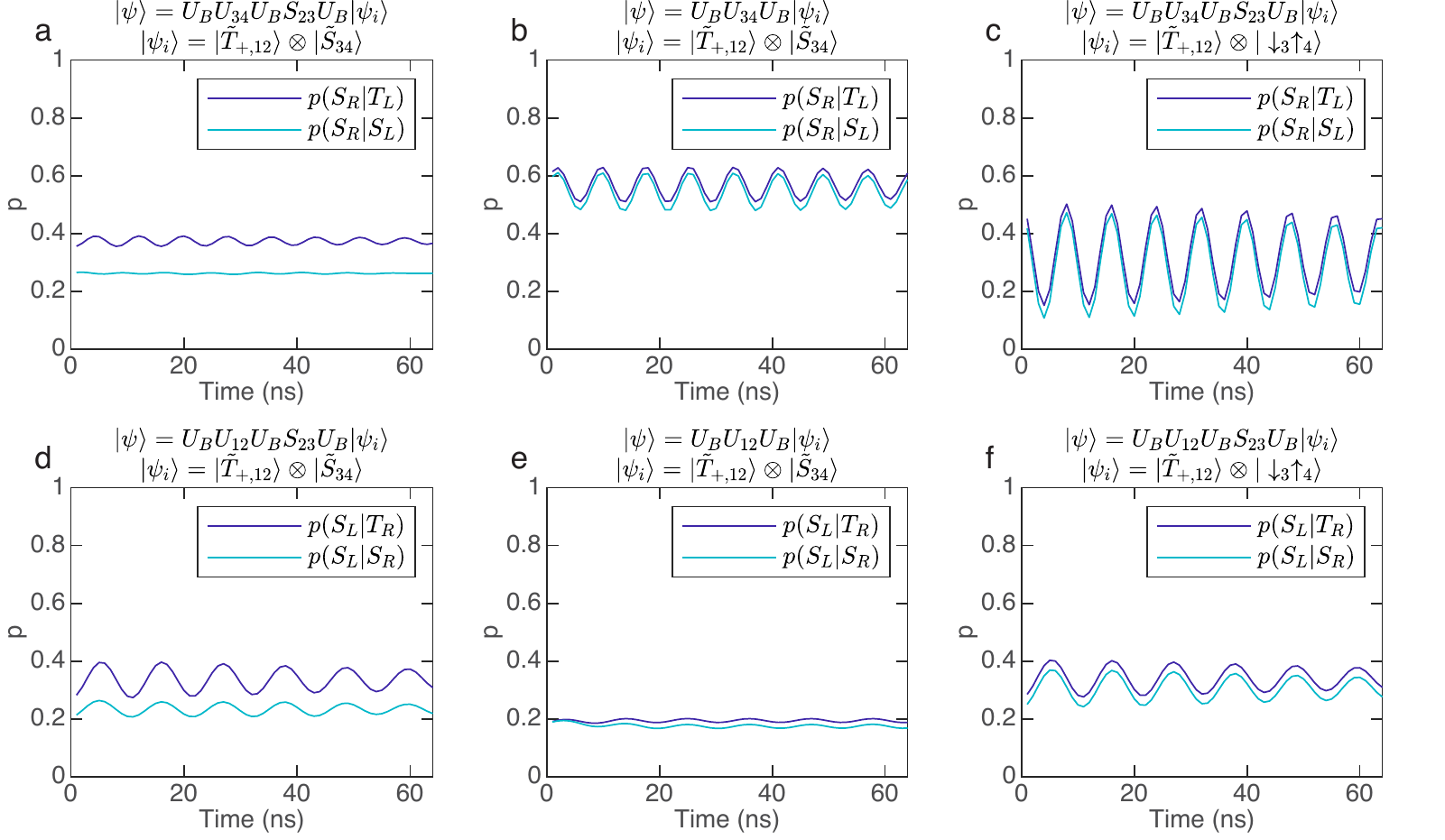}
	\caption{\label{xsim} Simulated variable-exchange experiments. Panels (a)-(f) present a simulation of the corresponding panel in Supplementary Fig.~\ref{fullExchange}. Above each panel is the operator sequence used in the simulation and the initial state used in the simulation. In panels (a)-(c), $p(S_R|S_L)$ indicates the right-side singlet probability given a singlet on the left, and $p(S_R|T_L)$ indicates the right-side singlet probability given a triplet on the left. In (d)-(f), $p(S_L|S_R)$ indicates the left-side singlet probability given a singlet on the right, and $p(S_L|T_R)$ indicates the left-side singlet probability given a triplet on the right. }
\end{figure*}

\begin{figure*}
	\centering
	\includegraphics{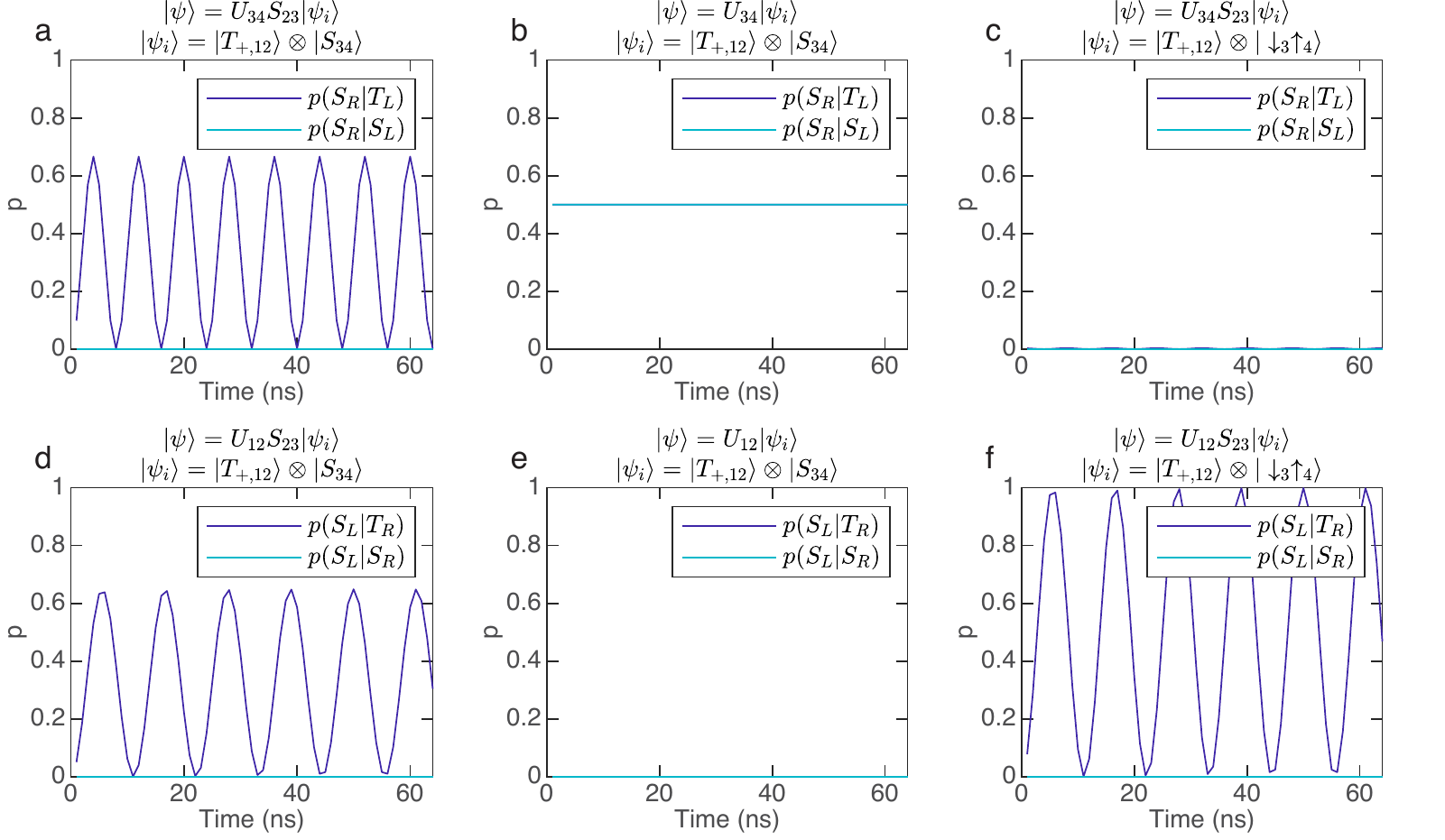}
	\caption{\label{xsimIdeal} Simulated variable-exchange experiments with no errors associated with pulses, hyperfine fluctuations, state preparation, or readout. Panels (a)-(f) present a simulation of the corresponding panel in Supplementary Fig.~\ref{fullExchange}. In panels (a)-(c), $p(S_R|S_L)$ indicates the right-side singlet probability given a singlet on the left, and $p(S_R|T_L)$ indicates the right-side singlet probability given a triplet on the left. In (d)-(f), $p(S_L|S_R)$ indicates the left-side singlet probability given a singlet on the right, and $p(S_L|T_R)$ indicates the left-side singlet probability given a triplet on the right.  The oscillations in (a) and (b) have visibility 2/3, as expected. In the case of (a), for example, there is a 1/3 chance that a triplet outcome on the left corresponds to a $\ket{T_0}$, in which case successful teleportation of the $\ket{\U}$ state occurs. The dominant mechanism that reduces the visibility of these oscillations in practice is the fluctuating sign of the hyperfine gradient. The absence of oscillations in (c) is a result of the fixed hyperfine gradient in this simulation. For example, if the gradient were such that the ground state of the right pair were $\ket{\U_3}\ket{\D_4}$, oscillations with unit visibility would be expected. Similarly, the unit-visibility oscillations in panel (f) occur because of the assumed ground orientation of qubits 3 and 4. The apparent conditional effect in this panel occurs because there is zero probability to measure a singlet on the right side, i.e., $p(S_R)=0$. }
\end{figure*}

\begin{figure*}
	\centering
	\includegraphics[width=6in]{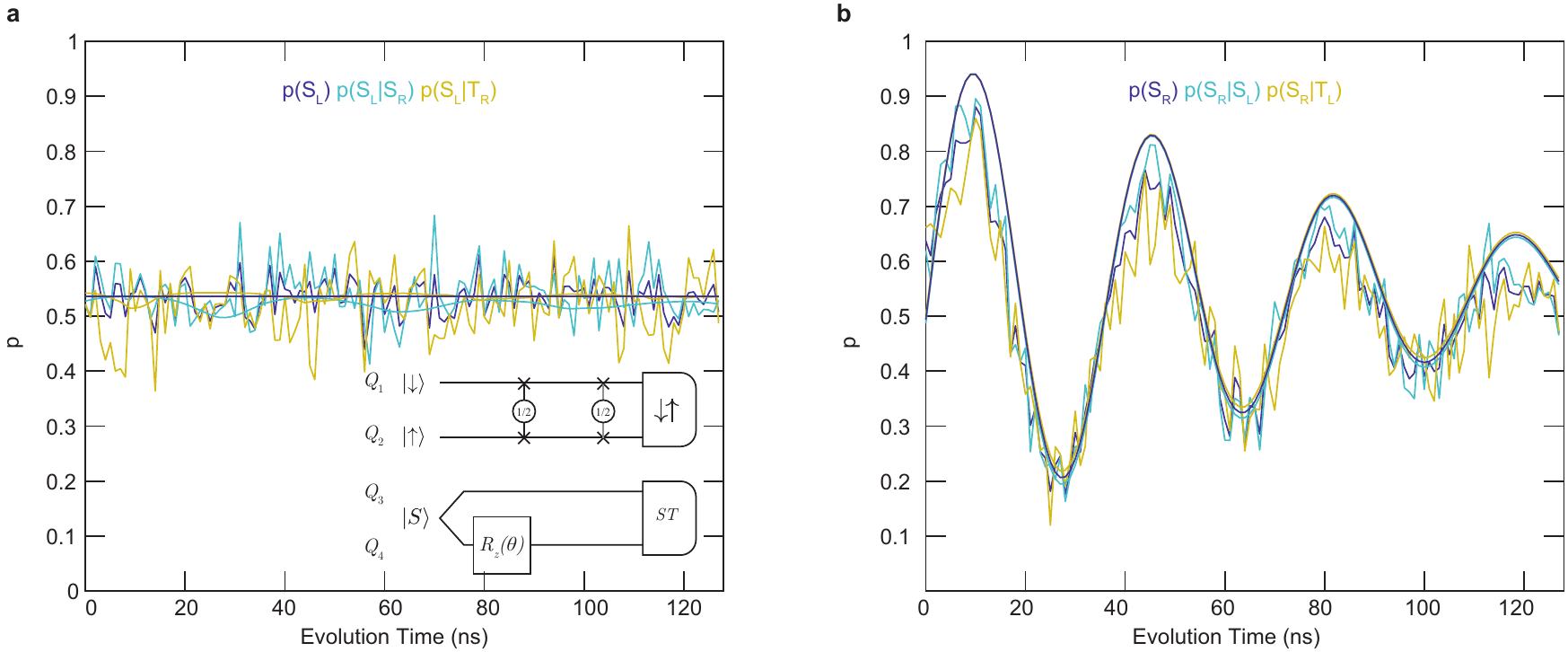}
	\caption{\label{swapcontrol} Control experiments for entanglement swapping and gate teleportation without the SWAP operation between qubits 2 and 3. To omit the SWAP gate, we enforce a wait for the same length of time with no voltage pulse. (a) Measured singlet probabilities on the left side. As before, simulations are overlaid in smooth lines of the same color. Inset: circuit diagram for the control experiment. (b) Measured right-side probabilities. Simulations are overlaid with smooth lines of the same color. The absence of any conditional effect in the left- and right-side measurements confirms that the behavior we observe in Fig.~\ref{entangle} is due to entanglement swapping and gate teleportation. These data were acquired in a separate run from the data in Fig.~\ref{entangle} in the main text. The data in panels (a) and (b) of this figure are from one repetition.}
\end{figure*}

\begin{figure*}
	\centering
	\includegraphics{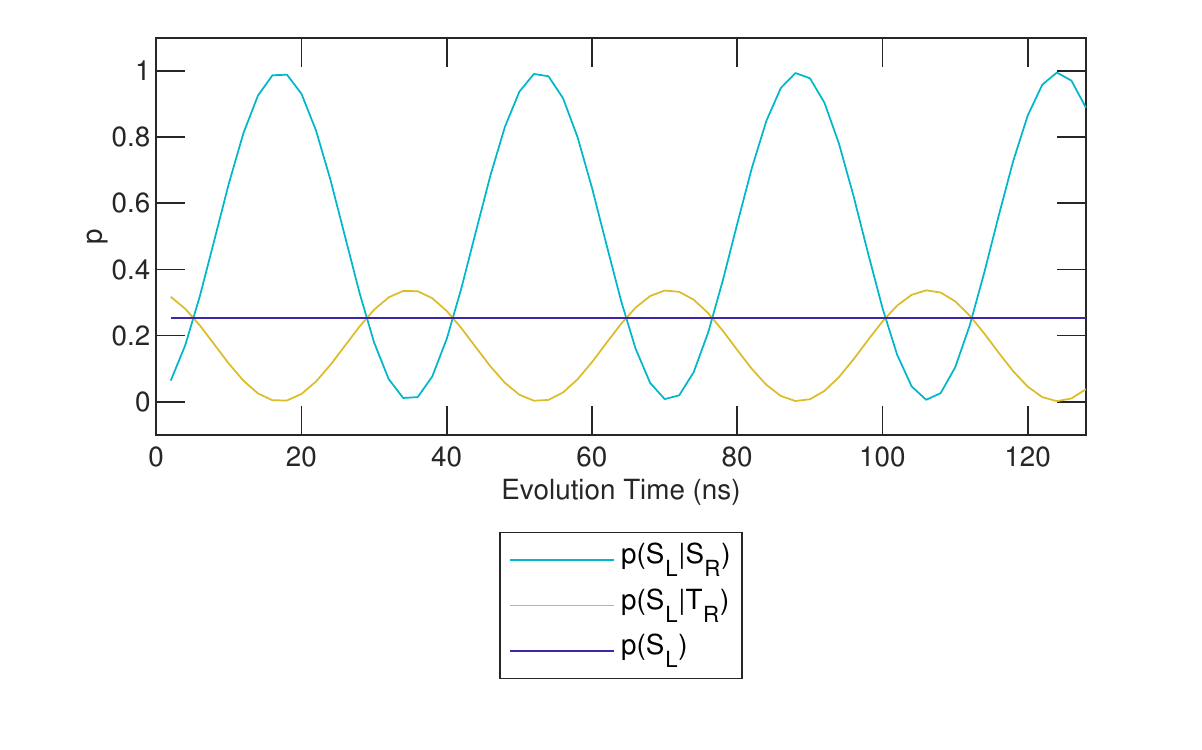}
	\caption{\label{entangleIdeal} Simulated entanglement swapping with no errors associated with pulses, hyperfine fluctuations, state preparation, or readout. The oscillations associated with $p(S_L|T_R)$ occur with visibility 1/3, because in 1/3 of the cases where we measure a triplet on the right pair, it is a $\ket{T_0}$, and successful teleportation has occurred. }
\end{figure*}

\begin{figure*}
	\centering
	\includegraphics[width=7in]{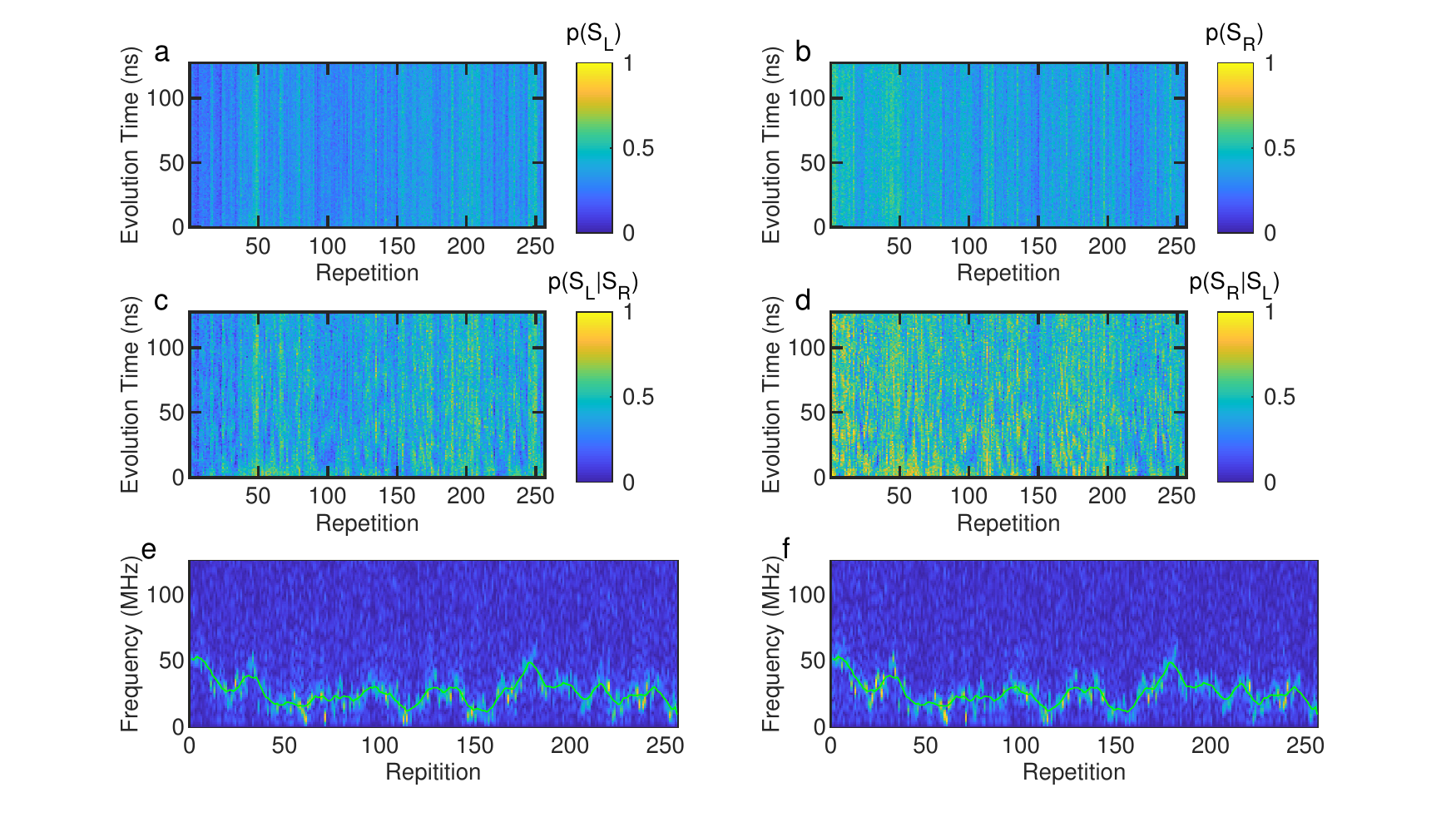}
	\caption{\label{fullEntangle} Conditional teleportation of entangled states and conditional gate teleportation. (a) Averaged singlet probability on the left pair of qubits $p(S_L)$. (b) Averaged singlet probability on the right pair of qubits $p(S_R)$. (c) Averaged singlet probability on the left pair of qubits given a singlet on the right pair $p(S_L|S_R)$. (d) Averaged singlet probability on the right pair of qubits given a singlet on the left pair $p(S_R|S_L)$. (e) Absolute value of the fast Fourier transform of the data in (c). The extracted peak frequency is overlaid in green. (f) Absolute value of the fast Fourier transform of the data in (d). The extracted peak frequency is overlaid in green.}
\end{figure*}

\begin{figure*}
	\centering
	\includegraphics[width=7in]{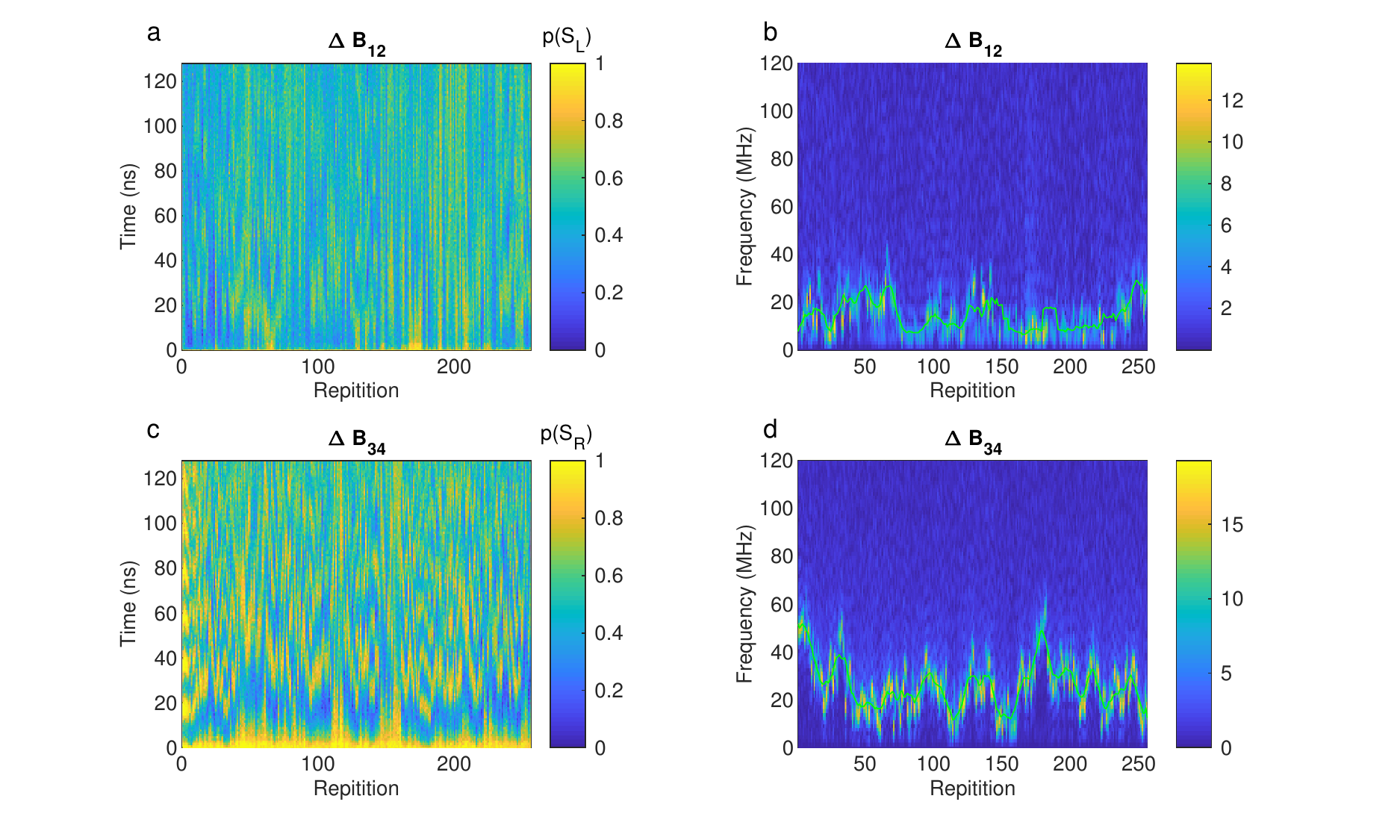}
	\caption{\label{dbz} Measurement of $\Delta B$. (a) Averaged measurements of $\Delta B_{12}$ oscillations ~\cite{Shulman2014}. Acquisition of each vertical line was interleaved with the data shown in Fig.~\ref{entangle}. (b) Absolute value of the fast Fourier transform of the data in (a), with the extracted frequency shown in green. (c) Averaged measurements of $\Delta B_{34}$ oscillations. Acquisition of these data were also interleaved with the data in Fig.~\ref{entangle}. (d) Absolute value of the fast Fourier transform of the data in (c), with the extracted frequency shown in green. }
\end{figure*}

\begin{figure*}
	\centering
	\includegraphics{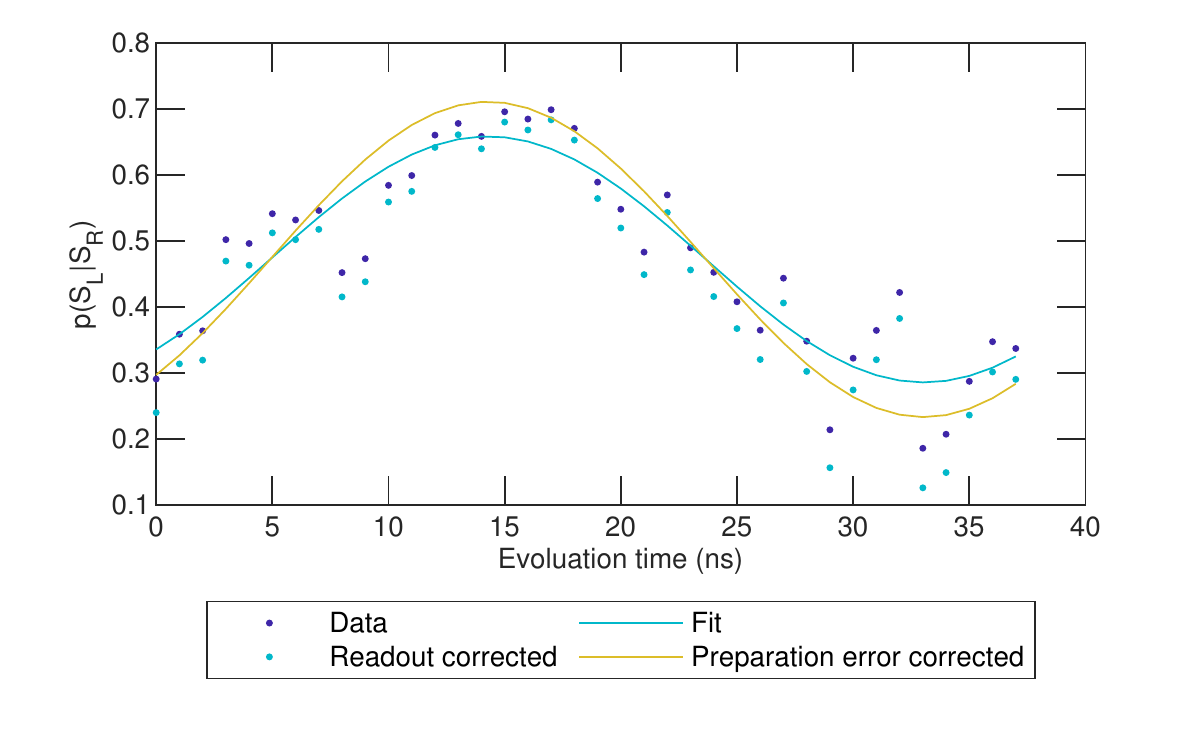}
	\caption{\label{swapFit} Experimental estimation of the maximum singlet teleportation probability. The dark blue dots are the data $p(S_L|S_R)$ from Fig.~\ref{entangle}(b). The light blue dots have been corrected for left-side readout errors. The light-blue line is a fit of the light-blue dots to a sinusoid. The gold line is the same sinusoid with the amplitude corrected for the right-side preparation error. The maximum singlet return probability is obtained at about 15 ns.}
\end{figure*}

\begin{figure*}
	\centering
	\includegraphics{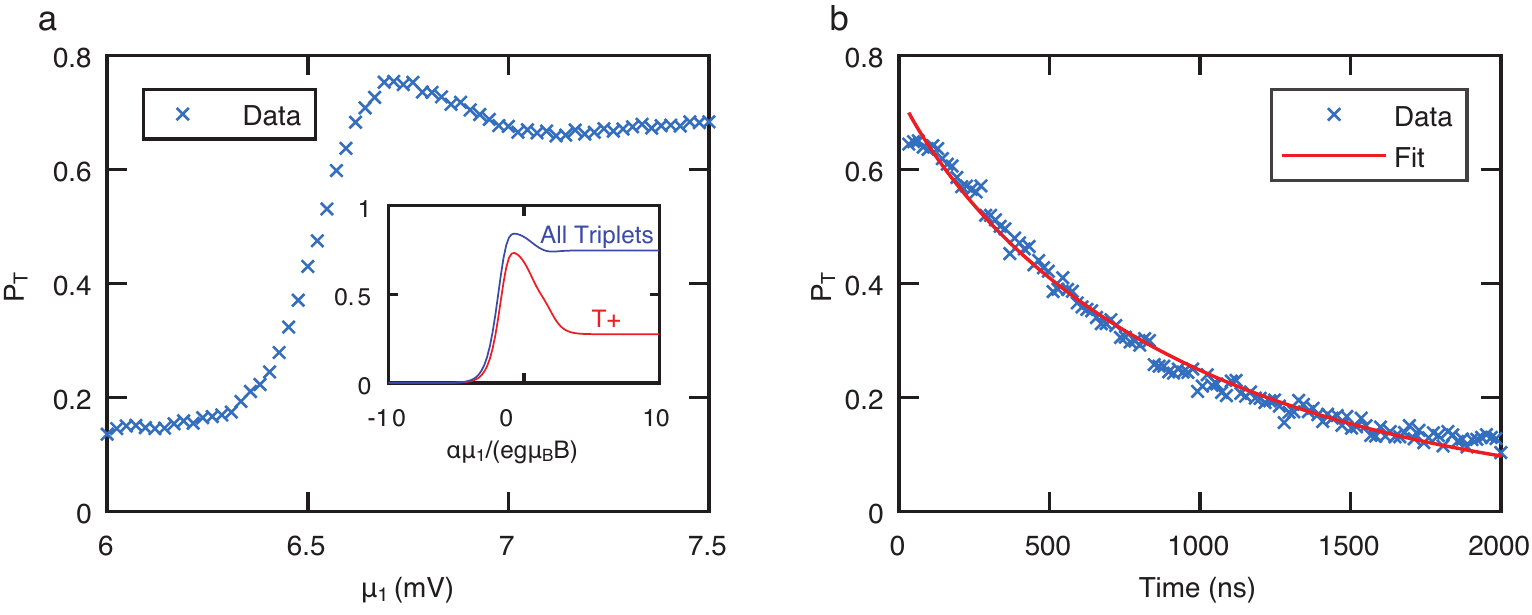}
	\caption{\label{Tp} State preparation fidelity. (a) Experimentally measured $\ket{T_+}$ loading curve, obtained by sweeping $\mu_1$, the electrochemical potential of dot 1, across the (1,1)-(2,1) charge transition, where $(m,n)$ indicates $m$ electrons in dot 1 and $n$ electrons in dot 2~\cite{Orona2018,Botzem2018}. The peak in the data indicates where the value of $\mu_1$ where the $\ket{T_+}$ loading probability is highest. Inset: Simulated results of the loading process. The blue simulation gives the expected triplet signal, and the red simulation is the probability of loading a $\ket{T_+}$. The simulation assumes a load time of 2 $\mu$s, a ramp time of $200$ ns, a temperature of 75 mK, and a magnetic field of 0.5 T. The $x$ axis represents the chemical potential $\alpha \mu_1$ in units of the magnetic field, where $\alpha$ is the effective lever-arm. (b) Triplet probability as a function of loading time. All experiments were conducted with a loading time of 2 $\mu$s. The blue points are data, and the red line is a fit to an exponential decay. These data are taken in the same tuning used to acquire all data in this work with one electron in each dot. }
\end{figure*}

\begin{figure*}
	\centering
	\includegraphics{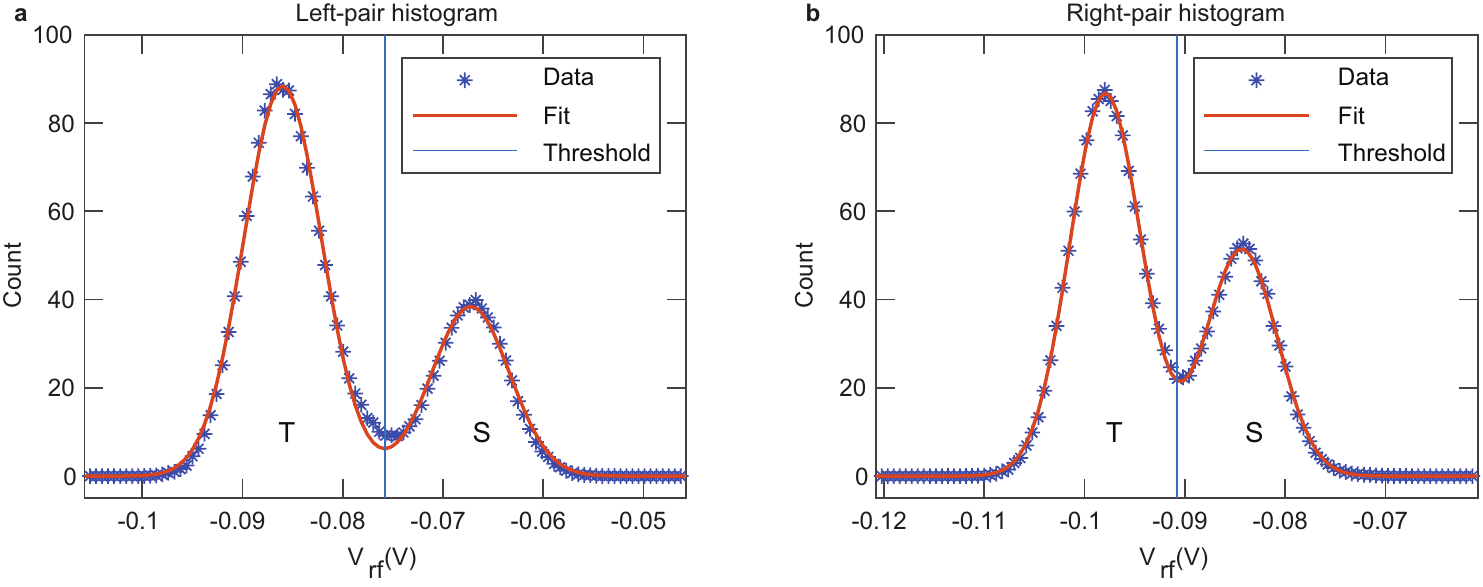}
	\caption{\label{readout} Readout fidelity. (a) Measurement histogram and fit for the left pair of qubits for the data shown in Fig.~\ref{teleport}(b). The extracted average fidelity for singlets and triplets is $0.93$. (b) Measurement histogram and fit for the right pair of qubits for the data shown in Fig.~\ref{teleport}(a). The extracted average fidelity for both singlets and triplets is $0.87$. In both panels, the red lines are fits to equations (1) and (2) of Ref.~\cite{Barthel2009}. In both panels, the $V_{RF}$ represents the raw voltage from our readout circuit. In both panels, the threshold which maximizes the visibility is indicated.}
\end{figure*}

\begin{figure*}
	\centering
	\includegraphics{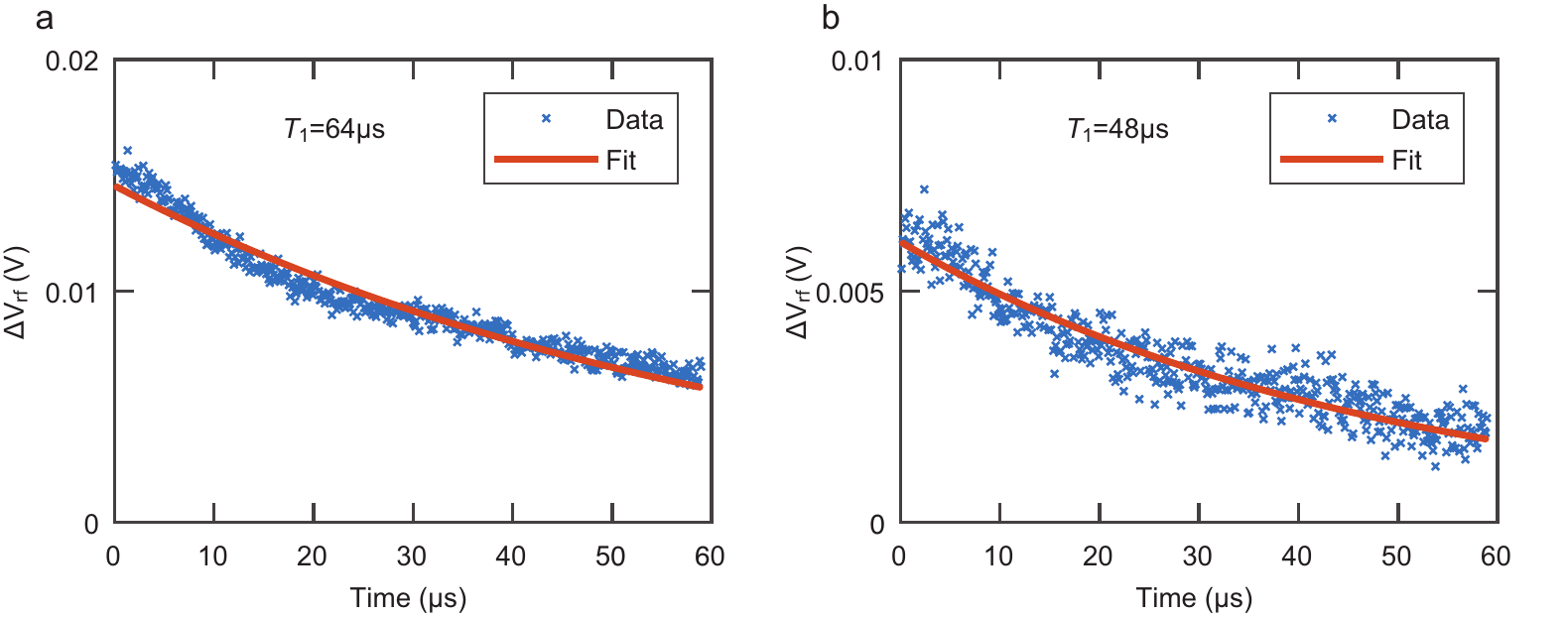}
	\caption{\label{T1} Relaxation during readout. (a) Data and fit showing the relaxation of the left pair of qubits during readout. (b) Data and relaxation of the right pair of qubits during readout. The data in both panels are obtained by subtracting the results of two measurements. The first involves preparing a mixed state and measuring it for 60 $\mu$s. Then, we prepare a singlet state and also measure it for 60 $\mu$s. We repeat this set of two measurements $10{,}000$ times. For each repetition, we record the entire 60 $\mu$s measurement record. We plot the difference of the two experiments, averaged across all repetitions, as a function of integration time from 0-60 $\mu$s. We fit this difference to a decaying exponential and extract the relaxation time $T_1$. The fits give relaxation times of 65 $\mu$s and 48 $\mu$s. These data are taken in the same tuning used to acquire all data in this work with one electron in each dot. In both panels, there appears to be a slight deviation from single-exponential decay. We attribute this deviation to the multiple relaxation pathways and rates involved in our shelving readout method.~\cite{Studenikin2012,Orona2018}.}
\end{figure*}

\begin{figure*}
	\includegraphics{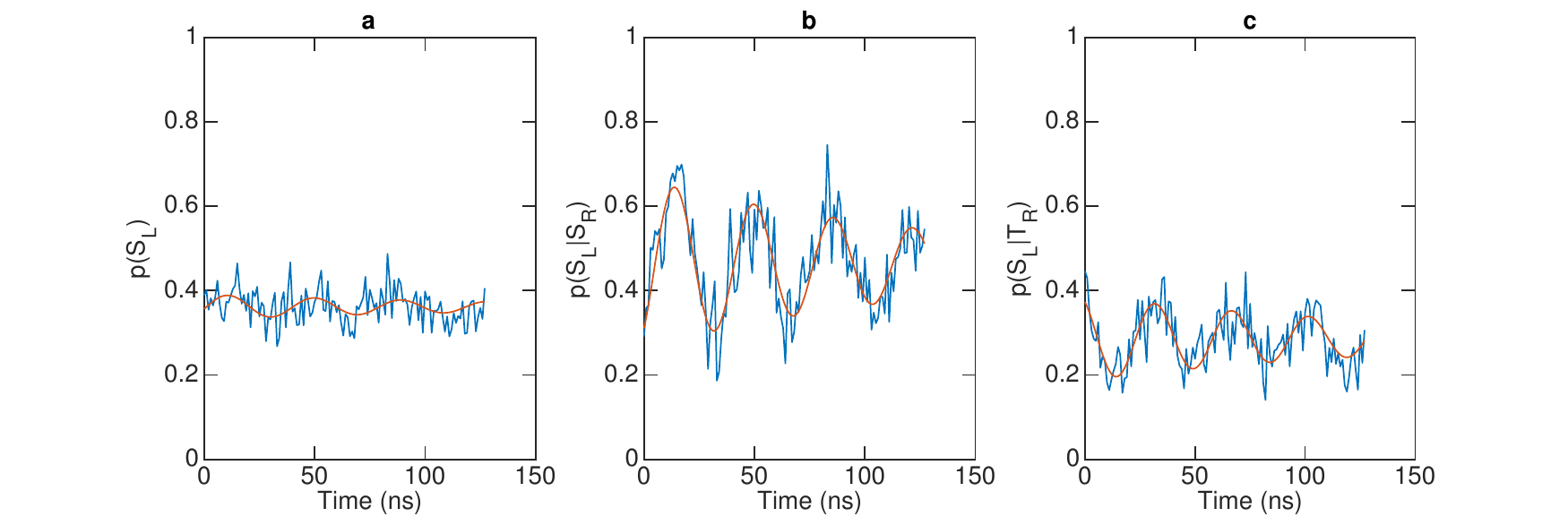}
	\caption{\label{pval} Data and fits. (a) $p(S_L)$ from Fig. 4b of the main text and fit. (b) $p(S_L|S_R)$ from Fig. 4b of the main text and fit. (c) $p(S_L|T_R)$ from Fig. 4b of the main text and fit. These data and fits are used to assess the probability that a classical teleportation strategy could reproduce the results of the entanglement-swap experiment.}
\end{figure*}

\begin{figure*}
	\includegraphics{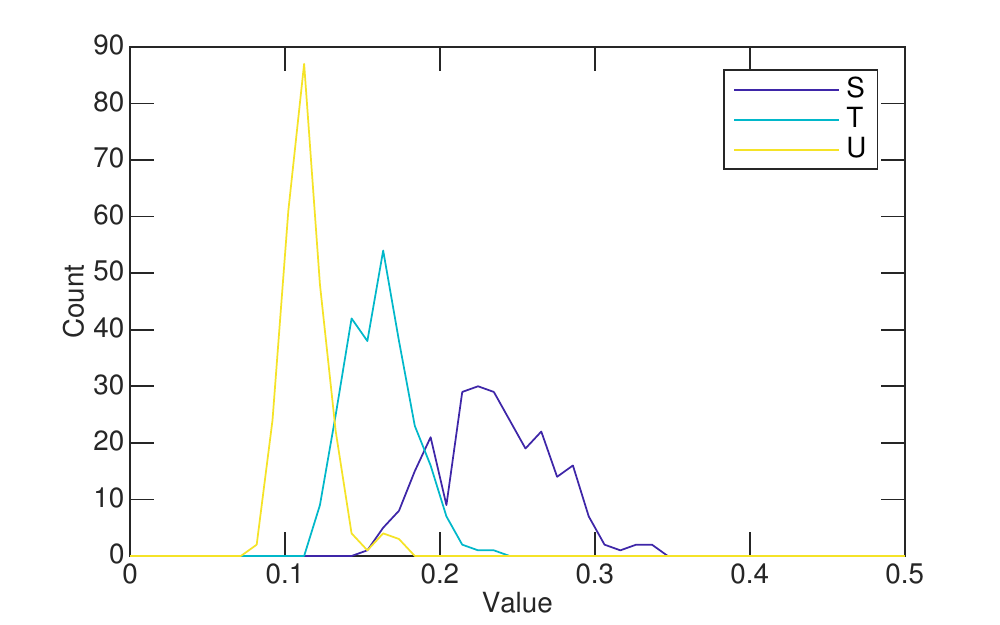}
	\caption{\label{pval2} Distributions of $S_i$, $T_i$, and $U_i$. The pronounced difference between these distributions indicates that a classical explanation for the entanglement-swap experiment is unlikely.}
\end{figure*}

\begin{figure*}
	\includegraphics{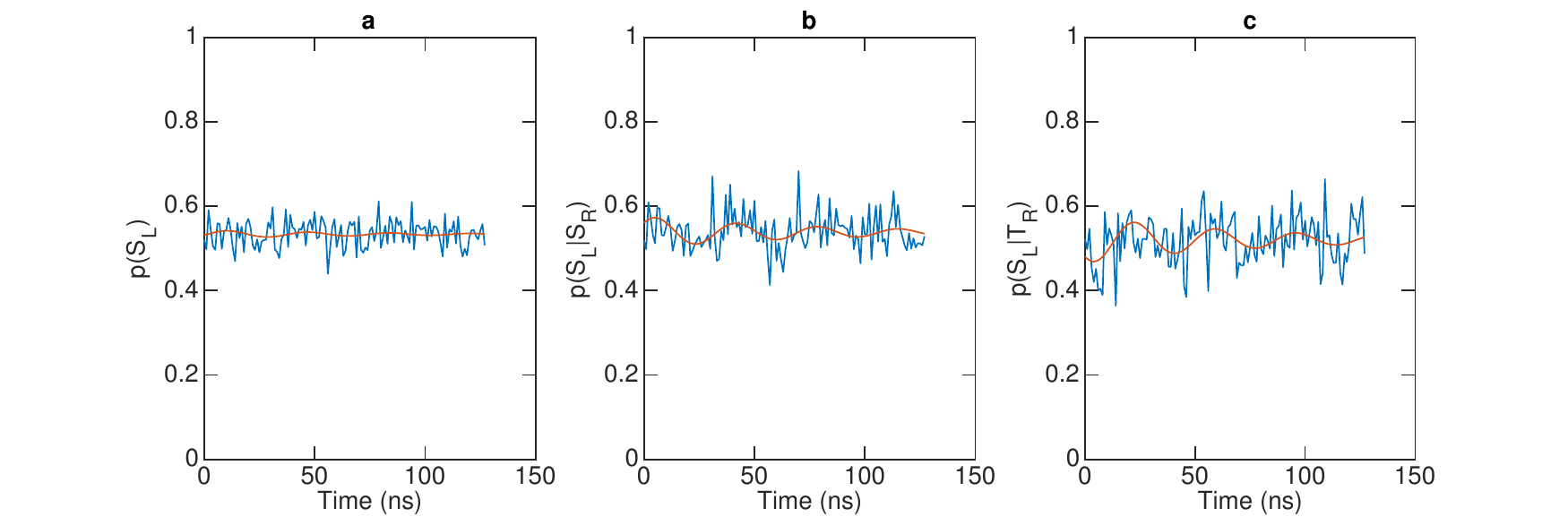}
	\caption{\label{pvalU} Data and fits for the control experiment. (a) $p(S_L)$ from Supplementary Fig. 5 and fit. (b) $p(S_L|S_R)$ from Supplementary Fig. 5. and fit. (c) $p(S_L|T_R)$ from Supplementary Fig. 5. and fit. These data and fits are used to assess the probability that spurious classical correlations could explain the entanglement-swap data.}
\end{figure*}

\clearpage
\section{Supplementary References}